\documentclass[10pt]{article}
\usepackage{amssymb,latexsym, amsmath}
\usepackage{amscd}

\makeatletter \@addtoreset{figure}{section}
\def\thefigure{\thesection.\@arabic\c@figure}
\def\fps@figure{h, t}
\@addtoreset{table}{bsection}
\def\thetable{\thesection.\@arabic\c@table}
\def\fps@table{h, t}
\@addtoreset{equation}{section}

\newtheorem{thm}{Theorem}[section]
\newtheorem{prop}[thm]{Proposition}
\newtheorem{lem}[thm]{Lemma}
\newtheorem{cor}[thm]{Corollary}
\newtheorem{remark}{Remark}[section]

\newcommand{\R}{\mathbb{ R}}

\newcommand{\g}{\mathfrak{ g}}

\DeclareMathOperator{\pr}{pr}
\DeclareMathOperator{\Span}{span}
\DeclareMathOperator{\ddim}{ddim}
\DeclareMathOperator{\tr}{tr}
\DeclareMathOperator{\diag}{diag}

\DeclareMathOperator{\Ad}{Ad}
\DeclareMathOperator{\rank}{rank}
\DeclareMathOperator{\ad}{ad}

\begin{document}
\title{Geodesic Flows and Neumann Systems on Stiefel Varieties.
Geometry and Integrability}

\author{Yuri N. Fedorov \and Bo\v zidar Jovanovi\' c}

\maketitle

\begin{abstract}
We study integrable geodesic flows on Stiefel varieties
$V_{n,r}=SO(n)/SO(n-r)$ given by the Euclidean, normal (standard),
Manakov-type, and Einstein metrics. We also consider natural
generalizations of the Neumann systems on $V_{n,r}$ with the above
metrics and proves their integrability in the non-commutative
sense by presenting compatible Poisson brackets on
$(T^*V_{n,r})/SO(r)$. Various reductions of the latter systems are
described, in particular, the generalized Neumann system on an
oriented Grassmannian $G_{n,r}$ and on a sphere $S^{n-1}$ in
presence of Yang-Mills fields or a magnetic monopole field.

Apart from the known Lax pair for generalized Neumann systems, an
alternative (dual) Lax pair is presented, which enables one to
formulate a generalization of the Chasles theorem relating the
trajectories of the systems and common linear spaces tangent to
confocal quadrics. Additionally, several extensions are
considered: the generalized Neumann system on the complex Stiefel
variety $W_{n,r}=U(n)/U(n-r)$, the matrix analogs of the double
and coupled Neumann systems.
\end{abstract}

\tableofcontents

\section{Introduction}
A Stiefel variety $V_{n,r}$ is the variety of $r$ ordered
orthogonal unit  vectors $(e_1,\dots, e_r)$ in the Euclidean space
$\R^n$ or, equivalently, the set of $n \times r$ matrices
$$
X=(e_1 \cdots e_r) \in M_{n,r}(\R)
$$
satisfying the condition
\begin{equation}
{X}^T {X}={\bf I}_r,
\label{cond_X*}
\end{equation}
where ${\bf I}_r$ is an $r\times r$ unit matrix (e.g, see
\cite{KN}). In particular, $V_{n,1}$ is a sphere $S^{n-1}$, while
both  $V_{n,n}$ and $V_{n,n-1}$ are diffeomorphic to $SO(n)$.

The integrable geodesic flows on $V_{n,2}$ and $V_{n,r}$,
$2<r<n-1$ are constructed in \cite{Th} and \cite{BJ1},
respectively. The geodesic flows were described in the homogeneous
space representation $SO(n)/SO(n-r)$ following a general approach
to the integrability of geodesic flows on homogeneous spaces
\cite{BJ1, BJ2, BJ3}.

In the first part of our paper (Sections 2, 3, 4) we study
geodesic flows in the redundant coordinates $X$ with  constraints
(\ref{cond_X*}) by using a Dirac approach and the corresponding
Poisson structure. This allows us to describe the flows in a quite
transparent way. Our main tool is the theorem on non-commutative
integrability of Hamiltonian systems (see \cite{N, MF2}) and for
various examples of integrable flows, we calculate the dimension
of invariant isotropic tori and give the matrix Lax
representations (Section 4).

For the Manakov-type metric on $V_{n,r}$ a geometric
interpretation of the motion in the form of the classical Chasles
theorem for the geodesic flow on an ellipsoid is given (Section
\ref{Yuri}). In addition, the complete integrability of the
geodesic flows of the $SO(n)$-invariant Einstein metrics
constructed by Jensen \cite{Je}, and Arvanitoyeorgos, Dzhepko and
Nikonorov \cite{ADN} are proved in Sections 4 and 6.

The second part of the paper (Sections 5, 6, 7, 8) considers
integrable Neumann flows on the Stiefel variety
$V_{n,r}=SO(n)/SO(n-r)$ and the oriented Grassmannian variety
$G_{n,r}=SO(n)/SO(r)\times SO(n-r)$, generalizing the famous
Neumann system on the sphere $S^{n-1}$. The latter is defined as a
natural mechanical system with the Hamiltonian (see \cite{Neum,
Moser, Moser2}):
\begin{equation}
H_{neum}=\frac12(p,p)+\frac12(Ae,e), \qquad A=\diag(a_1,\dots,a_n) ,
\label{sphere_neum}
\end{equation}
where the cotangent bundle $T^*S^{n-1}$ is realized as a
submanifold of $\R^{2n}\{e,p\}$ given by the constraints
$(e,e)=1$,  $(e,p)=0$. This system, together with the Jacobi
problem on the geodesic flow on an ellipsoid, provides one of the
basic and most beautiful examples of application of algebraic
geometric tools to integrable systems (e.g, see \cite{Fl, Mum,
AHP, AHH}). It is also directly related to many other integrable
models (e.g., see \cite{Moser2, Ve, Knorr1, Kuz, FeJo}).

The Neumann systems on $V_{n,r}$ which we consider have the
kinetic energy of $SO(n)$-invariant metrics described in Section 3
and the potential function
$$
V=\frac12\tr(X^TAX)=\frac12\sum_{i=1}^r (e_i,Ae_i).
$$

Two matrix Lax representations are presented. The first, a "big"
one, given by Theorem \ref{Neumann_LA} is closely related to the
symmetric Clebsch--Perelomov rigid body problem \cite{Pe}. For
$r=1$, it was given by Moser in \cite{Moser} and for $r>1$ and the
case of the Manakov type submersion metrics by Reyman and
Semenov--Tian-Shanski \cite{RS} within the framework of the $R$-matrix
method.  Note that for $r>1$ this Lax pair does not define a
Neumann system on $V_{n,r}$ uniquely and does not provide a non-commutative set of integrals,
necessary for the integrability.

In contrast, the second (dual, or "small") Lax pairs, given by
Theorem \ref{DUAL} are equivalent to the Neumann systems with the
Euclidean and normal metrics up to an action of a finite discrete group.
For the Neumann system with the Euclidean metric, the small Lax
pair was first given in the unpublished manuscript \cite{Kap89}.

In Sections 6 and 7 we give a detailed proof of the
non-commutative integrability of the considered Neumann flows by
using the Bolsinov completeness condition for a set of Casimir
functions of the pencil of compatible Poisson brackets (see
\cite{Bo}). We also indicate an integrable generalization of the
Neumann system on Grassmannians with a quartic potential.

In Section \ref{Yuri} we propose a geometric interpretation of the integrals of the
Neumann systems on $V_{n,r}$ obtained from the dual Lax
representation. Our geometric model generalizes the celebrated
Chasles theorem adopted by Moser for the case $r=1$ (see, e.g, Theorem 4.10
in \cite{Mum}).

Magnetic Neumann flows in the rank two case (on $G_{n,2}$ and
$V_{n,2}$) as well as the motion of a particle on a sphere
$S^{n-1}$ under the influence of a Yang--Mills field are presented
in Appendix 1. Finally in Appendix 2, we briefly consider the rank
$r$ double and coupled Neumann flows, as well as an extension of
the Neumann system onto a complex Stiefel variety
$W_{n,r}=U(n)/U(n-r)$.

The geodesic flows and Neumann systems considered in this paper
are written in a form appropriate for their integrable
discretizations, which we describe in a separate paper
\cite{FeJo2}.

\section{Hamiltonian Flows on Stiefel Varieties}
\label{first}
\paragraph{Stiefel varieties.}
As it was mentioned in Introduction, a Stiefel variety $V_{n,r}$
is the variety of $r$ ordered  orthogonal unit vectors
$(e_1,\dots, e_r)$ in the Euclidean space $\left({\mathbb R}^n,
(\cdot,\cdot)\right)$,  or, equivalently, the set of $n \times r$
matrices satisfying constraints (\ref{cond_X*}). Thus $V_{n,r}$ is
a smooth subvariety of dimension $N=rn-r(r+1)/2$ in the space of
$n\times r$ real matrices $M_{n,r}(\R)=\R^{nr}$ and the components
of ${X}$ are redundant coordinates on it.

The left $SO(n)$ action on $V_{n,r}$ ($X\mapsto RX$, $R\in SO(n)$)
is transitive, hence $V_{n,r}$ can be realized as a homogeneous
space of the Lie group $SO(n)$ as well. If fix the orthonormal
base in $\R^n$ 
\begin{equation}
E_1=(1,0,,0,\dots,0)^T,\quad E_2=(0,1,0,\dots,0)^T,\, \dots, \; \;
E_n=(0,0,0,\dots,1)^T  \label{base}
\end{equation}
and take the point $X_0=(E_1,\dots,E_r)\in V_{n,r}$, then the
orthogonal transformation fixing $X_0$ (relative to the above
basis of $\R^n$) must have the form
\begin{equation}
\begin{pmatrix}
{\bf I}_r & 0 \\
0 & B
\end{pmatrix}, \quad B\in SO(n-r).
\label{hat}
\end{equation}
Since the isotropy group of $X_0$ is isomorphic to $SO(n-r)$, the
variety $V_{n,r}$ can be identified with $SO(n)/SO(n-r)$.

\paragraph{The Poisson structure.}
The tangent bundle $TV_{n,r}$ is the set of pairs $({X}, \dot
{X})$ subject to the constraints
\begin{equation} \label{cond_X}
{X}^T {X}={\bf I}_r, \quad {X}^T
\dot{X} + \dot{X}^T {X} =0.
\end{equation}
On the other hand, the cotangent bundle $T^*V_{n,r}$ can be
realized as the set of pairs of $n\times r$ matrices $({X},{P})$
that satisfy the constraints
\begin{equation}
\label{cond_XP}{X}^T {X}
={\bf I}_r, \quad {X}^T {P} + {P}^T {X} =0\, .
\end{equation}
The latter give $r(r+1)$ independent scalar constraints
\begin{equation}
F_{ij}=(e_i,e_j)-\delta_{ij}=0, \quad G_{ij} =
(e_i,p_j)+(e_j,p_i)=0, \quad 1 \le i \le j \le r,\label{scalar}
\end{equation}
where $p_j$ is the $j$-th column of the matrix $P$. This
realization of $T^*V_{n,r}$ is motivated by description of the
geodesic flows of the Euclidean and normal metric on $V_{n,r}$
given below, however there are other natural realizations of
$T^*V_{n,r}$ (see Section \ref{flows}).

The canonical symplectic structure $\omega$ on $T^*V_{n,r}$ is the
restriction of the canonical 2-form in the ambient space
$T^*M_{n,r}(\R)$,
$$
\omega_0=\sum_{i=1}^n \sum_{s=1}^r d{p}_{s}^i\wedge \, d{e}_{s}^i\, .
$$

For our purposes it is convenient to work with the redundant
variables $(X,P)$. The canonical Poisson structure on $T^*V_{n,r}$
can then be described by using the Dirac construction \cite{AKN,
Dirac, Moser}. Namely, let $\{\cdot,\cdot\}_0$ be the canonical
Poisson bracket on $\R^{2nr}$
$$
\{f_1,f_2\}_0=\sum_{i=1}^r
\left(
\left( \frac{\partial f_1}{\partial e_i}, \frac{\partial f_2}{\partial p_i} \right)
-\left( \frac{\partial f_1}{\partial p_i},\frac{\partial f_2}{\partial e_i}\right)\right)
$$
and $C_{i,j}$ be the inverse of the matrix $\{F_i,F_j\}_0$, $i,j=1,\dots,r(r+1)$,
where, for the sake of simplicity, we denoted constraints (\ref{scalar}) by $F_i=0$, $i=1,\dots,r(r+1)$.
Then the Dirac bracket is given by
\begin{equation}
\{f_1,f_2\}=\{f_1,f_2\}_0
+\sum_{i,j} \{F_i,f_1\}_0 C_{i,j} \{F_j,f_2\}_0.
\label{Dirac_bracket}
\end{equation}
The subvariety $T^*V_{n,r}$ appears as a symplectic leaf of the
Dirac bracket and the restriction of $\{f_1,f_2\}$ to $T^*V_{n,r}$
depends only on the restriction of $f_1$ and $f_2$ to
$T^*V_{n,r}$.

The Hamiltonian equation
$$
\dot f=\{f,H\}
$$
can be also described by the using Lagrange multipliers. We shall write them in the matrix form:
\begin{equation}
\begin{aligned}
&\dot X=\frac{\partial H}{\partial P} - X\Pi, \\
&\dot P=- \frac{\partial H}{\partial X} + X\Lambda + P\Pi\, ,
\end{aligned}
\label{dot_XP}
\end{equation}
where $\Lambda$ and $\Pi$ are $r\times r$ symmetric matrix Lagrange multipliers
uniquely determined from the condition for the
trajectory $(X(t),P(t))$ to satisfy constraints (\ref{cond_XP}).

\paragraph{Momentum mappings.}
The Lie group $SO(n)$ naturally acts on $T^*V_{n,r}$ by left
multiplication:
\begin{equation}
R\cdot (X,P)=(RX,RP), \quad R\in SO(n) .
\label{left_action}
\end{equation}

Below we use the well known identification of $\Lambda^2\R^n$ with
a subset of $so(n)$: $x\wedge y=x\otimes y-y\otimes x=x\cdot
y^T-y\cdot x^T$, $x,y\in\R^n$. Also, $\langle\cdot,\cdot\rangle$
is proportional to the Killing metric on $so(n)$: \begin{equation}
\langle \xi_1,\xi_2\rangle=-\frac12 \tr(\xi_1 \xi_2),
\label{killing}\end{equation} $\xi_1, \xi_2\in so(n)$. By the use
of the scalar product (\ref{killing}) we identify $so(n)$ and
$so(n)^*$.

\begin{prop} \label{Phi}
The left $SO(n)$-action (\ref{left_action}) is Hamiltonian. The
equivariant momentum mapping $\Phi: T^*V_{n,r} \to so(n)^*\cong
so(n)$ is given by
\begin{equation}
\Phi(X,P)=\sum_{i=1}^r p_i \wedge e_i,
\label{momentum_map}
\end{equation}
or, in the matrix form
$
\Phi(X,P)=P X^T-X P^T.
$
\end{prop}

\noindent{\it Proof.}
The left $SO(n)$-action (\ref{left_action})
on $(T^*M_{n,r}(\R), \{\cdot,\cdot\}_0)$ is Hamiltonian
with the momentum map (\ref{momentum_map}).
The constraint functions $X^T X$ and $X^T P + P^T X$ are $SO(n)$-invariant.
Therefore
\begin{equation}
\{\Phi_\xi,F_{i}\}_0=0,  \quad i=1,\dots,r(r+1),
\label{zero}
\end{equation}
where $\Phi_\xi (X,P)=\langle \Phi(X,P),\xi\rangle$ is the Hamiltonian function
of the action of the one-parameter subgroup $\{\exp(s\xi), s\in\R\}$.

In view of the definition of the Dirac bracket (\ref{Dirac_bracket}) and
(\ref{zero}), the Hamiltonian flows of $\Phi_\xi$ on
$T^*V_{n,r}$ with respect to the brackets $\{\cdot,\cdot\}_0$ and
$\{\cdot,\cdot\}$ coincide. This proves the proposition.
\hfill$\Box$

\medskip

Together with a left $SO(n)$-action, we also have
the natural right $SO(r)$-action:
\begin{equation} \label{right_action}
(X,P)\cdot Q=(X Q, P Q), \quad Q\in SO(r).
\end{equation}

Following similar lines, one can prove

\begin{prop}
The right $SO(r)$-action (\ref{right_action}) is Hamiltonian. The
equivariant momentum mapping $\Psi: T^*V_{n,r} \to so(r)^* \cong
so(r)$ is given by
\begin{equation} \Psi(X,P)=X^T P - P^T X\, .
\label{momentum_right_map} \end{equation}
\end{prop}

The momentum mappings $\Phi$ and $\Psi$ are Poisson with respect to
the (+) and (-) Lie-Poisson brackets on $so(n)$ and $so(r)$:
\begin{eqnarray*}
&&\{h_1\circ \Phi(X,P),h_2\circ \Phi(X,P)\}=
\{h_1(\mu),h_2(\mu)\}^+_{so(n)}, \quad \mu=\Phi(X,P), \\
&&\{f_1\circ \Psi(X,P),f_2\circ \Psi(X,P)\}=
\{f_1(\eta),f_2(\eta)\}^-_{so(r)}, \quad \eta=\Psi(X,P),
\end{eqnarray*}
where
\begin{eqnarray*}
&&\{h_1(\mu),h_2(\mu)\}^+_{so(n)}=\langle \mu, [\nabla h_1(\mu),\nabla h_2(\mu)]\rangle,
\quad h_1,h_2: so(n)\to \R, \\
&&\{f_1(\eta),f_2(\eta)\}^-_{so(r)}=-\langle \eta, [\nabla f_1(\eta),\nabla f_2(\eta)]\rangle,
\quad f_1,f_2: so(r)\to \R
\end{eqnarray*}
and where the brackets $\langle\cdot,\cdot\rangle$ denote the
scalar product (\ref{killing}) on $so(n)$ and $so(r)$,
respectively.

\paragraph{The algebra of $SO(n)$-invariant functions.}
Let $C^\infty(T^*V_{n,r})^{SO(n)}$ be the algebra of
$SO(n)$-invariant functions on $T^*V_{n,r}$. Since $SO(n)$ acts in
a Hamiltonian way on $T^*V_{n,r}$, $C^\infty(T^*V_{n,r})^{SO(n)}$
is closed under the Poisson bracket.

Let $X_0=(E_1,\dots,E_r)$. The $SO(n)$-invariant functions, via
restrictions, are in one-to-one correspondence with the
$SO(n-r)$-invariant functions on $T^*_{X_0} V_{n,r}$.

We can write $P_0\in T^*_{X_0} V_{n,r}$ as a block matrix
\begin{equation}\label{P-neutral}
 P_0=
\begin{pmatrix}
P_1 \\
P_2
\end{pmatrix},
\end{equation}
where $P_1$ and $P_2$ are $r\times r$ and $(n-r)\times r$
matrixes, respectively. Then $P_0$ satisfies constraints
(\ref{cond_XP}) at $X=X_0$ if $P_1^T=-P_1$. Also, the
$SO(n-r)$-action on $T^*_{X_0} V_{n,r}$ is given by
\begin{equation}
P_1\longmapsto P_1, \qquad P_2 \longmapsto  B\cdot P_2, \qquad B\in SO(n-r). \label{SO(n-r)-action}
\end{equation}

\begin{lem}
The maximal number of functionally independent $SO(n)$-invariant
functions, i.e., the differential dimension of
$C^\infty(T^*V_{n,r})^{SO(n)}$, equals
\begin{equation}
\ddim C^\infty(T^*V_{n,r})^{SO(n)}= \left\{\begin{array}{cc} \dim
V_{n,r}-\dim SO(n-r), & n\le 2r
\\ \dim V_{n,r}-\dim SO(n-r)+\dim SO(n-2r), & n>2r .
\end{array}
\right. \label{ddim}
\end{equation}
\end{lem}

\noindent{\it Proof.} The differential dimension of
 $C^\infty(T^*V_{n,r})^{SO(n)}$ is just the codimension of the
generic $SO(n-r)$ orbit in $T^*_{X_0} V_{n,r}$. The dimension of
the orbit $SO(n-r)\cdot P_0$ is
$$
\dim SO(n-r)-\dim SO(n-r)_{P_0},
$$ where $SO(n-r)_{P_0}$ is the isotropy group of $P_0$. Since
$SO(n-r)_{P_0} =\{{\bf I}_{n-r}\}$ for $n\le 2r$, for a generic
$P_0$ we get the first relation in \eqref{ddim}. Further, in the
case $n>2r$, we can take $ P_0=(E_{r+1},E_{r+2},\dots,E_{2r}). $
Then $SO(n-r)_{P_0} = SO(n-2r)$. \hfill$\Box$

\medskip

It can easily be verified that the restrictions of the $SO(n)$-invariant functions
\begin{equation}
\Psi_{ij}=(e_i,p_j)-(e_j,p_i), \quad (P^TP)_{ij}=(p_i,p_j), \quad i,j=1,\dots,r
\label{SO(n)-inv}
\end{equation}
to $T^*_{X_0}V_{n,r}$ define the generic orbits of the action
(\ref{SO(n-r)-action}). In particular, we get the following simple
statement.

\begin{lem}\label{LEMA}
If a smooth function $f\in C^\infty(T^*V_{n,r})$ Poisson commutes
with functions (\ref{SO(n)-inv}), then it Poisson commutes with all
$SO(n)$-invariant functions on $T^*V_{n,r}$.
\end{lem}

The Poisson bracket on $C^\infty(T^*V_{n,r})^{SO(n)}$ can be
described as follows. The restriction of momentum mapping $\Phi$
to $T^*_{X_0} V_{n,r}$ establish the isomorphism
\begin{equation}
T^*_{X_0} V_{n,r} \cong \mathfrak v, \label{iso}
\end{equation}
where $\mathfrak v$ is the orthogonal complement of $so(n-r)$ in
$so(n)$. Within identification (\ref{iso}), the $SO(n-r)$-action
(\ref{SO(n-r)-action}) corresponds to the adjoint $SO(n-r)$-action
on $\mathfrak v$ and the Poisson bracket on
$C^\infty(T^*V_{n,r})^{SO(n)}$ corresponds  to the Poisson bracket
$$
\{\tilde f_1(\xi),\tilde f_2(\xi)\}_\mathfrak v=
-\langle \xi, [\nabla \tilde f_1(\xi),\nabla \tilde f_2(\xi)]\rangle.
$$
on the algebra $C^{\infty}(\mathfrak v)^{SO(n-r)}$ of $SO(n-r)$-adjoint invariants on
$\mathfrak v$ (see Thimm \cite{Th}).

Recall that the set of commuting $SO(n)$-invariant functions
$\mathfrak A$ is complete if it contains maximal possible number
of independent functions, that is (see \cite{BJ1, BJ3}),
\begin{equation*}
\ddim \mathfrak A =\dim V_{n,r}-l,
\end{equation*}
where $2l$ is the dimension of a generic adjoint orbit in
$$
\Phi(T^*V_{n,r})=\Ad_{SO(n)} \Phi(T^*_{X_0} V_{n,r}).
$$
Note that, for $n \le 2r+1$, the generic orbit in
$\Phi(T^*V_{n,r})$ is regular, while it is singular otherwise. One
can prove the following relations.

\begin{lem}\label{2l}
$$
2l=\left\{\begin{array}{cc} \frac{n(n-1)}{2}-
\left[\frac{n}{2}\right], & n\le 2r+1, \\ 2r(n-r-1), & n>2r+1.
\end{array}
\right.
$$
\end{lem}

Now consider the chain of subalgebras
${so}(n-r+1)\subset {so}(n-r+2) \subset \dots \subset{so}(n)$,
where a matrix $\xi\in so(n-r+i)$ is included in $so(n)$ as a block matrix
\begin{equation}
\begin{pmatrix}
{0} & 0 \\
0 & \xi
\end{pmatrix}.
\label{hat2}
\end{equation}
Let $\mathfrak A_i$ be the algebra of invariants on $so(n-r+i)$
considered as a polynomials on $so(n)$ and restricted to
$\mathfrak v$. Then
\begin{equation}\label{com}
\mathfrak A=\mathfrak A_1+\dots+\mathfrak A_{r} \end{equation} is
a complete polynomial commutative subset of $C^{\infty}(\mathfrak
v)^{SO(n-r)}$ (see again \cite{BJ1, BJ3}). Other complete
commutative sets of $SO(n)$-invariant functions are given in
\cite{DGJ} and Theorem \ref{CIF} below.

\section{Geodesic Flows}
\label{flows}

\paragraph{$SO(n)$-invariant metrics.}
An $SO(n)$-invariant metric $g$ on $V_{n,r}$ can be specified by a
positive definite, $SO(n-r)$-invariant scalar products $g_0$ at the point
$X_0$ as follows:
$$
g(R \cdot \eta_1 , R \cdot \eta_2)_X = g_0(\eta_1,\eta_2), \quad
\eta_1,\eta_2\in T_{X_0} V_{n,r}, \quad X= R\cdot X_0.
$$

Equivalently, an $SO(n)$-invariant metric can be defined by
using an $SO(n)$-invariant Hamiltonian function $H(X,P)$, which is
quadratic in momenta and positive definite on $T^*V_{n,r}$.

There are two natural $SO(n)$-invariant metrics on the Stiefel
variety $V_{n,r}$: the metric induced from the embedding of
$V_{n,r}$ in the Euclidean space $M_{n,r}(\R)$ and the normal
metric induced from a bi-invariant metric on the Lie group
$SO(n)$ (see below).

Concerning geometrical significance, one should also mention
invariant Einstein metrics (see \cite{Be}) constructed in \cite{Je}, \cite{ADN}.

\paragraph{The Euclidean metric.}
The Euclidean metric in $M_{n,r}(\R)$ is given by the Lagrangian function
$$
L_E (X,\dot X)=\frac12 \tr(\dot X^T \dot X)=
\frac12\sum_{i=1}^r  (\dot e_i,\dot e_i).
$$
The Legendre transformation
\begin{equation}
P=\frac{\partial L_E}{\partial \dot X}=\dot X
\label{leg}
\end{equation}
yields the Hamiltonian function
\begin{equation}
H_E(X,P)=\frac12\tr(P^T P)=\sum_{i=1}^r \frac12 (p_i,p_i)
\label{Euclidean}
\end{equation}
defined on the cotangent bundle $T^*M_{n,r}(\R)$.

We shall refer to the restriction of the above metric to $V_{n,r}$
as the {\it Euclidean metric}, which will be denoted by $ds^2_E$.
The geodesic flow is described by the Euler--Lagrange equations
with multipliers
$$
\frac{d}{dt}\frac{\partial L_E}{\partial \dot X}=\frac{\partial L_E}{\partial X} +
X\Lambda \quad \Longleftrightarrow \quad
\ddot X=X\Lambda, \quad \Lambda=-\dot X^T \dot X \, ,
$$
where the symmetric matrix $\Lambda$ is uniquely determined from the
condition for the trajectory $X(t)$ to satisfy the constraints ${X}^T{X}={\bf I}_r$.

Taking into account constraints (\ref{cond_X}) and the Legendre
transformation (\ref{leg}), we see that the cotangent bundle
$T^*V_{n,r}$ can be represent as a submanifold of $T^*M_{n,r}(\R)$
given by (\ref{cond_XP}). The corresponding
Hamiltonian flow of $H_E(X,P)$ with respect to the Dirac bracket is
\begin{equation}
\begin{aligned}
&\dot X=P, \\
&\dot P=-XP^T P .
\end{aligned}
\label{Euclidean_XP}
\end{equation}

\paragraph{Submersion metrics.}
Let $\bar g(\cdot,\cdot)$ be a right-invariant metric on $SO(n)$.
The subgroup $SO(n-r)$ acts freely on $SO(n)$ by isometries (right
action). There is a $\bar g$-orthogonal decomposition of $T_R
SO(n)$
$$
\mathcal{SO}(n-r)_R+\mathcal H_R=T_R SO(n), \quad R\in SO(n) ,
$$
where $\mathcal{SO}(n-r)_R$ is the tangent space to the fibre $R\cdot
SO(n-r)$. By definition, the submersion metric $g(\cdot,\cdot)$ on
$SO(n)/SO(n-r)$ is given by
$$
g(\xi_1,\xi_2)_{\rho(R)} =
\bar g(\bar \xi_1,\bar \xi_2)_R,
\quad \xi_i \in T_{\rho(R)}(SO(n)/SO(n-r)), \; \bar \xi_i \in \mathcal H_R, \;
\xi_i=d\rho(\bar \xi_i),
$$
where $\rho: SO(n)\to SO(n)/SO(n-r)$ is the canonical projection
(e.g., see \cite{Be}). The vectors in $\mathcal{SO}(n-r)_R$ and
$\mathcal H_R$ are called vertical and  horizontal respectively.

The Hamiltonian of a right invariant metric on $SO(n)$ can be
written in the form $h\circ\bar\Phi$, where $h$ is a positive
definite quadratic form on $so(n)$ and $\bar\Phi: T^*SO(n)\to
so(n)$ is the momentum mapping of the natural left $SO(n)$-action.
It follows that the class of submersion metrics on $SO(n)/SO(n-r)$
is given by Hamiltonian functions of the form $h\circ \Phi$, where
now $\Phi$ is the momentum mapping of a natural left
$SO(n)$-action on $T^*(SO(n)/SO(n-r))$ (e.g., see \cite{Br}).

The above observation helps us to write down the
Hamiltonians and the geodesic flows of the submersion metrics in
the representation of the cotangent bundle $T^*V_{n,r}$ given by the
constraints (\ref{cond_XP}). The Hamiltonians are 
\begin{equation}
H_\mathbb A (X,P)=h_\mathbb A\circ \Phi=
\frac12\langle \mathbb A\Phi,\Phi\rangle=
-\frac14\tr\left(\mathbb A(P X^T  - X P^T)(P X^T -X P^T )\right),
\label{submersion_metric}
\end{equation}
where $h_\mathbb A(\xi)=\frac12\langle \mathbb A\xi,\xi\rangle$, $\xi\in so(n)$
and $\mathbb A: so(n)\to so(n)$ are positive definite operators.

\begin{prop}
The equations of the submersion metrics geodesic flow generated
by (\ref{submersion_metric}) are
\begin{equation}
\begin{aligned}
& \dot X=\mathbb A(\Phi)\cdot X, \\
& \dot P=\mathbb A(\Phi)\cdot P.
\end{aligned}\label{dot_XP2}
\end{equation}
\end{prop}

\noindent{\it Proof.} By using the chain rule $dH=dh_\mathbb
A\circ d\Phi$, one gets expressions (\ref{dot_XP2}) for the
derivatives of the Hamiltonian (\ref{submersion_metric}) with
respect to $P$ and $-X$. Further, it can easily be
verified that then the Lagrange matrix multipliers $\Lambda$ and
$\Pi$ in (\ref{dot_XP}) are zero. \hfill$\Box$

\begin{remark}{\rm
In particular, for $r=n$, the system (\ref{dot_XP2}) describes the
right-invariant geodesic flow on the Lie group $SO(n)$. The
symmetric form of the equations differs from the symmetric
representation of the rigid body equations given in \cite{BCMR}.
}\end{remark}

\paragraph{The normal metric.}
If $\bar g(\cdot,\cdot)$ is a bi-invariant metric
induced by the scalar product (\ref{killing}), then the
submersion metric is called the {\it normal metric}. It is
proportional to the {\it standard metric} induced by the negative
Killing form \cite{Be}) on $SO(n)/SO(n-r)$. We shall denote the
normal metric by $ds^2_0$. Contrary to a generic submersion
metric, the normal metric is also $SO(n)$-invariant. The corresponding Hamiltonian has the form
\begin{equation}
H_0(X,P)=\frac12 \langle \Phi,\Phi \rangle=\frac12\tr(P^TP)-\frac12\tr((X^T P)^2),
\label{normal_metric}
\end{equation}
and its geodesic flow is given by
\begin{eqnarray}
&& \dot X=\Phi\cdot X=P-X P^T X, \label{dot_X3}\\
&& \dot P=\Phi\cdot P=-XP^T P+ PX^T P. \label{dot_P3}
\end{eqnarray}

Under the conditions (\ref{cond_XP}), the relation (\ref{dot_X3}) can be uniquely inverted and one gets
\begin{equation}
P=\dot X- \frac12 XX^T \dot X  . \label{P(dot_X)}
\end{equation}
Relations (\ref{dot_X3}), (\ref{P(dot_X)}) give
identification of $TV_{n,r}$ and $T^* V_{n,r}$ by means of the
normal metric. Therefore, the Lagrangian function for the metric
$ds^2_0$ is
\begin{equation}\label{normal_metric_lagrangian}
L_0(X,\dot X)=\frac12 \langle \Phi_0,\Phi_0\rangle ,
\end{equation}
where $\Phi_0(X,\dot X)=\Phi(X,P(X,\dot X))$:
\begin{equation}
\Phi_0(X,\dot X)=\dot X X^T-X\dot X^T-\frac 12 {X} [ {X}^T \dot  {X}-  \dot{X}^T {X}] {X}^T .
\label{Phi_0}
\end{equation}

For a {\it right} $SO(n)$-action (and
therefore with opposite signs in the equations), the relation (\ref{Phi_0}) is established in \cite{FeJo} by
studying nonholonomic LR systems on the Lie group $SO(n)$. Note
that the realization of $T^*V_{n,r}$ via (\ref{cond_XP}) is also
natural if we consider the flow given by the normal metric.
Namely, defining momenta by the Legendre transformation
$P={\partial L_0}/{\partial \dot X},$ from the constraints
(\ref{cond_X}) we get the conditions (\ref{cond_XP}).

\paragraph{Einstein metrics.}
The momentum mappings $\Phi$ and $\Psi$ are invariant under the
$SO(r)$ and $SO(n)$ actions, respectively. Therefore, the
Hamiltonians of the form
\begin{equation}
H_\kappa(X,P)=\frac12 \langle \Phi,\Phi \rangle+\frac{\kappa}{2}
\langle \Psi,\Psi
\rangle=\frac12\tr(P^TP)-\left(\frac12+\kappa\right)\tr((X^T P)^2)
\label{EM}
\end{equation}
is $SO(n)\times SO(r)$ invariant. Within the class of the metrics
$ds^2_\kappa$ determined by the Hamiltonian functions (\ref{EM}) there is
the normal metric ($\kappa=0$) and the Euclidean metric
($\kappa=-1/2$). Moreover, for $r=2$ there is a unique value of $\kappa$, while for
$r>2$ there are exactly two values, such that
$ds^2_\kappa$ is an Einstein metric (see \cite{Je, ADN}). Following
\cite{ADN}, we refer to these metrics as {\it the Jensen} metrics.

Further, in \cite{ADN}, new examples of the Einstein metrics are
obtained within the class of metrics that we shall describe below.

Consider the Lie subalgebra
$$
so(r_1)\oplus so(r_2)\oplus \dots\oplus so(r_k) \subset so(r),
\quad r_1+r_2+\dots+ r_k=r,
$$
naturally embedded into $so(r)$. Define the Hamiltonian  $H_K$ via
\begin{equation}
H_K=\frac12 \langle \Phi,\Phi \rangle+K\circ \Psi , \label{EM2}
\end{equation}
where the quadratic function $K$ is
\begin{equation}
K(\xi)= \frac{\kappa_0}2\langle \xi,\xi\rangle+
\frac{\kappa_1}2\langle \xi_1,\xi_1 \rangle + \dots+
\frac{\kappa_k}2\langle \xi_k,\xi_k \rangle. \label{EM3}
\end{equation}
Here $\xi_i$ are orthogonal projections (with respect to the
$so(r)$-Killing metric) to $so(r_i)\subset so(r)$.

In  \cite{ADN} Arvanitoyeorgos,  Dzhepko, and Nikonorov proved that if
$r_1=r_2=\dots=r_k$, $\kappa_1=\kappa_2=\dots=\kappa_k$,
$k>1$, $n-r>r_1 \ge 3$,
than among the metrics defined by Hamiltonians (\ref{EM2}) there
are four Einstein metrics (two of them, with $\kappa_1=0$, are the
Jensen metrics).

\paragraph{$SO(r)$-invariant metrics.}
Let $A$ be a symmetric, positive definite $n\times n$ matrix.
The geodesic flows on $V_{n,r}$ with Lagrangians of the form
$$
L_A(X,\dot X)=\frac12\tr(\dot X^T A \dot X)=
\frac12 \sum_{i=1}^r (\dot e_i, A \dot e_i),
$$
were studied in \cite{BCS} from the point of view of the variational and optimal control problems.
The Lagrangian $L_A$ can be considered on the whole space $M_{n\times r}(\R)$,
where the Legendre transformation
\begin{equation}
P=\frac{\partial L_A}{\partial \dot X}=A\dot X,
\label{leg2}
\end{equation}
gives the following Hamiltonian on the cotangent bundle $T^*M_{n,r}(\R)$
$$
H_A(X,P)=\frac12\tr(P^T A^{-1} P)=\frac12 \sum_{i=1}^r (p_i,A^{-1}p_i) .
$$
>From (\ref{cond_X}) and (\ref{leg2}) we conclude that the
cotangent bundle $T^*V_{n,r}$ can be represented as a submanifold of
$T^*M_{n,r}(\R)$ given by the equations
\begin{equation}
\label{cond_XP2}
{X}^T {X}={\bf I}_r, \quad {X}^T A^{-1}{P} + {P}^T A^{-1} {X} =0.
\end{equation}
Again, one defines the Dirac bracket with respect to the constraints (\ref{cond_XP2}).

Let $ds^2_A$ be the metric defined by the Lagrangian $L_A$ and $\{\cdot,\cdot\}_A$ be
the new Dirac bracket. Then the geodesic flow of $ds^2_A$ can be described by the Hamilton equations
$$
\dot f=\{f,H_A (X,P) \}_A
$$
restricted to the symplectic leaf (\ref{cond_XP2}). The corresponding matrix form of the flow is
\begin{equation}
\begin{aligned}
&\dot X=A^{-1}P, \\
&\dot P=X\Lambda ,
\end{aligned}\label{A_XP}
\end{equation}
where $\Lambda$ is a symmetric $r\times r$ matrix uniquely determined from the condition
$$
\Lambda X^T A^{-1} X + X^T A^{-1} X\Lambda+2 P^T A^{-2} P=0.
$$

Note that for $r=1$, the metric $ds^2_A$ is a standard metric on
the ellipsoid $(x,A^{-1}x)=1$, while for $r$ it is a
right-invariant Manakov rigid body metric on $SO(n)$ (see
\cite{Moser}). Furthermore, $ds^2_A$ is right $SO(r)$-invariant
and, via submersion, induces a metric on the oriented Grassmannian
variety $G_{n,r}$ (see Section \ref{grassmannian}). It would be
interesting to prove integrability of the corresponding geodesic
flows.

\section{Integrability of Geodesic Flows} \label{geod_flows}

\paragraph{The normal metric.}
As shown in \cite{BJ1, BJ2}, the geodesic flows of the normal
metrics $ds^2_0$ on the homogeneous spaces $G/H$ of compact Lie
groups $G$ are completely integrable in the non-commutative sense.
The proof is based on the following geometrical observation. Let
$(M,G,\Phi)$ be a Hamiltonian $G$-space with an equivariant
momentum mapping $\Phi: M\to \g^*$, where $G$ is a compact group.
Consider the following two natural sets of functions on $M$: the
functions obtained by pulling-back the algebra $C^\infty(\mathfrak
g^*)$ by the moment map and the set of $G$-invariant functions
$C^\infty(M)^G$. They are closed under the Poisson bracket and
according to the Noether theorem $\{\Phi^*(C^\infty(\mathfrak
g^*)),C^\infty(M)^G\}=0$ \cite{GS}. Moreover,
$\Phi^*(C^\infty(\mathfrak g^*))+C^\infty(M)^G$ is a complete set
of functions \cite{BJ2}. That is, any Hamiltonian system with
those integrals is non-commutatively integrable \cite{N, MF2}. In
particular, consider the case when $(M,G,\Phi)$ is a cotangent
bundle $T^*(G/H)$ with a natural $G$ action. Since the Hamiltonian
$H_0$ of the normal metric is of the form $h\circ\Phi$, where $h$
is an invariant quadratic polynomial on $\g^*$, the function $H_0$
Poisson commutes with all $G$-invariant functions (the Noether
theorem), as well as with all the functions in
$\Phi^*(C^\infty(\g^*))$ (the mapping $h \mapsto h \circ\Phi$ is a
morphism of Poisson structures). Therefore, the flow of $H_0$ is
non-commutatively integrable.


Let $2\, l$ be the dimension of  a generic orbit in
$\Phi(T^*V_{n,r})$ (see Lemma \ref{2l}).

\begin{thm} \textup{\cite{BJ1, BJ2}}\label{bols-jov}
The geodesic flow of the normal metric (\ref{dot_X3}),
(\ref{dot_P3}) is completely integrable in the non-commutative
sense. The complete algebra of first integrals is
$$
\Phi^*(C^\infty(so(n))+C^\infty(T^*V_{n,r})^{SO(n)}.
$$
The generic motions of the system are quasi-periodic over the isotropic tori
of dimension
$$
\delta_0 = 2\dim V_{n,r}-\ddim C^\infty(T^*V_{n,r})^{SO(n)}-2l.
$$
\end{thm}

Recall that the number of functionally independent
$SO(n)$-invariant functions is given by (\ref{ddim}). Thus we have
$$
\delta_0=\bigg\{\begin{matrix}
r\,, & n \ge 2r \\
\dim V_{n,r}+\dim SO(n-r)-\dim SO(n)+\rank SO(n)\,, & n<2r
\end{matrix}.
$$

It is interesting that for $n\ge 2r$ the dimension of the invariant tori
coincides with the dimension of invariant tori of geodesic
flows of normal metrics on the corresponding Grassmannian
manifolds $G_{n,r}$.

\paragraph{Manakov metrics.}
The above construction has a natural generalization to a class of
geodesic flows of submersion metrics given by the Hamiltonians
(\ref{submersion_metric}), such that the corresponding Euler
equations on $so(n)$:
\begin{equation}
\dot f=\{f,h_\mathbb A\}^+_{so(n)} \quad \Longleftrightarrow \quad
\dot \xi=[\nabla h_\mathbb A(\xi),\xi]=[\mathbb A \xi,\xi], \quad \xi\in so(n)
\label{Euler-so(n)}
\end{equation}
are completely integrable. For example, choose the Manakov operator
\begin{equation}
\mathbb A(E_i\wedge E_j)=\frac{b_i-b_j}{a_i-a_j} E_i\wedge E_j,
\quad 1\le i<j \le n, \quad \mathrm{i.e.,} \quad \mathbb A(\xi)=\ad_A^{-1}\ad_B\xi,
\label{Manakov}
\end{equation}
where all the eigenvalues of $A=\diag(a_1,\dots,a_n)$ and
$B=\diag(b_1,\dots,b_n)$ are distinct and $\mathbb A$ is positive
definite. Then the Euler equations (\ref{Euler-so(n)}) are
completely integrable (\cite{Ma, Fe}). Moreover, for generic $A$,
among the integrals $\tr(\xi+\lambda A)^k$ one can always find a
complete  set of commuting integrals $h_1,\dots,h_l$ on a generic
adjoint orbit in  $\Phi(T^*V_{n,r})$ (see Brailov \cite{Br} and
Bolsinov \cite{Bo}). Therefore, according to Theorem 2.2 in \cite{BJ2}, we have

\begin{thm}
The geodesic flow of the submersion metric (\ref{dot_XP2})
with the metric given by the Manakov operator (\ref{Manakov}) is completely integrable in the non-commutative
sense with a complete set of polynomial integrals (\ref{SO(n)-inv}) and
\begin{equation} \label{man}
\tr\left(\lambda A+\sum_{i=1}^r p_i\wedge e_i\right)^k, \quad k=1,\dots,n, \quad \lambda\in\R.
\end{equation}
The Generic motion of the system is quasi-periodic over the isotropic tori
of dimension
$$
\delta=2\dim V_{n,r}-\ddim C^\infty(T^*V_{n,r})^{SO(n)}-l.
$$
\end{thm}

As in the case of geodesic flows of normal metrics, for $n \ge
2r$ the dimension of generic invariant tori is simply $\delta=r(n-r)$.

\begin{remark}{\rm
The non-commutative integrability implies the usual commutative
integrability, at least by means of smooth commuting integrals
\cite{BJ2}. For the case of the  geodesic flows considered above,
the commuting integrals can be taken to be the polynomials (\ref{com}) and (\ref{man}).}
\end{remark}

Like in the case of the right-invariant metric on $SO(n)$, the
Manakov metric on $V_{n,r}$ possess a matrix Lax pair.

\begin{thm}
Equations (\ref{dot_XP2}) with the metric given by the Manakov
operator (\ref{Manakov}) imply the matrix equation with a spectral parameter $\lambda$
\begin{gather*}
\frac{d}{dt} \mathcal L_{man}(\lambda)=[\mathcal A_{man}(\lambda),\mathcal L_{man}(\lambda)] \, , \\
\mathcal L_{man}(\lambda)=\Phi+\lambda A, \quad \mathcal A_{man}(\lambda)=\mathbb A(\Phi)+\lambda B\,.
\end{gather*}
\end{thm}

\paragraph{The dual Lax pair for the Manakov flows.}
Consider the Manakov operator (\ref{Manakov}) for $B=A^2$. Then
\begin{equation}
\mathbb A(\xi)=A\xi+\xi A ,
\label{Manakov*}
\end{equation}
and equations (\ref{dot_XP2}) become
\begin{equation}
\begin{aligned}
& \dot X=A(PX^T-XP^T)X+(PX^T-XP^T)AX, \\
& \dot P=A(PX^T-XP^T)P+(PX^T-XP^T)AP.
\end{aligned}\label{dot_XP3*}
\end{equation}

\begin{thm} \label{man-dual}
Up to the action of a discrete group $\mathbb Z_2^n$ generated by
reflections with respect to the coordinate hyperplanes in $\R^n$,
\begin{gather} \label{reflections}
(X,P)\longmapsto (S_i X,S_i P), \qquad i=1,\dots,n, \\
S_i(x_1,\dots,x_n)=(y_1,\dots,y_n), \quad y_j=x_j, \quad j\ne i,
\quad y_i=-x_i, \notag
\end{gather}
the geodesic flow (\ref{dot_XP3*}) is equivalent to the matrix equation
\begin{equation}
\frac{d}{dt} \mathcal L^*_{man}(\lambda)=[\mathcal L^*_{man}(\lambda),\mathcal A^*_{man}(\lambda)]
\label{MAN-LA}
\end{equation}
with a spectral parameter $\lambda$ and $2r\times 2r$ matrices
\begin{eqnarray}
&& \mathcal L^*_{man}(\lambda)=\begin{pmatrix}
-X^T(\lambda \mathbf{I}_n-A)^{-1} P & -X^T(\lambda \mathbf{I}_n-A)^{-1} X \\
P^T(\lambda\mathbf{I}_n-A)^{-1} P & P^T(\lambda \mathbf{I}_n-A)^{-1} X
\end{pmatrix}, \label{MAN-LA2} \\
&& \mathcal A^*_{man}(\lambda)=\begin{pmatrix}
X^T(A+\lambda\mathbf{I}_n)^{-1} P & X^T(A+\lambda \mathbf{I}_n)^{-1} X \\
-P^T(A+\lambda\mathbf{I}_n)^{-1} P & -P^T(A+\lambda\mathbf{I}_n)^{-1} X
\end{pmatrix}.
\label{LA-MAN3}
\end{eqnarray}
\end{thm}

Note that after imposing the condition $X^TP=0$, equations
(\ref{dot_XP3*}) formally coincide with the equations describing
rank $r$ solutions of the Manakov system on $so(n)$ (see
\cite{Fe2}).

\paragraph{The Euclidean and Jensen's metrics.}
Since the Hamiltonian (\ref{EM}) is $SO(n)\times SO(r)$-invariant,
we can apply the general construction used in Theorem
\ref{bols-jov} with respect to the $SO(n)\times SO(r)$-action (see
\cite{BJ2}). Let $C^\infty(T^*V_{n,r})^{SO(n)\times SO(r)}$ be the
algebra of $SO(n)\times SO(r)$-invariant functions on
$T^*V_{n,r}$.

\begin{thm} \label{ET}
The geodesic flows of Jensen's metrics $ds^2_\kappa$ with
the Hamiltonian functions (\ref{EM}) are completely integrable in
the non-commutative sense. The complete algebra of first integrals is
\begin{equation}\label{so(n)+so(r)}
\Phi^*(C^\infty(so(n))+\Psi^*(C^\infty(so(r))+C^\infty(T^*V_{n,r})^{SO(n)\times
SO(r)}.
\end{equation}
In particular, the geodesic flow  (\ref{Euclidean_XP}) of the
Euclidean metric is completely integrable.
\end{thm}

The complete commutative set of polynomials $\mathfrak F$ within
$C^\infty(T^*V_{n,r})^{SO(n)\times SO(r)}$ as well as the
integrability of the geodesic flows with Hamiltonians (\ref{EM2})
will be given below (see \eqref{FF} in Section 6, Theorem
\ref{CIF} and Corollary \ref{nikinorov}).

\begin{remark}{\rm
Both the geodesic flows of the Euclidean and the normal metric
share the isotropic foliation defined by integrals
\eqref{so(n)+so(r)}, but do not share the isotropic foliation
defined in Theorem \ref{bols-jov}. Namely, the  straightforward
calculations show that functions $(p_i,p_j)$ in (\ref{SO(n)-inv})
do not Poisson commute with $\langle\Psi,\Psi\rangle$, and,
therefore, the algebra of $SO(n)$-invariant functions is not
conserved along the geodesic flow of the Euclidean metric (see
Lemma \ref{LEMA}). }\end{remark}

Here we note the following characterization of $SO(n)\times
SO(r)$-invariant metrics.

\begin{prop}
If the metric $ds^2$ on $V_{n,r}$ is $SO(n)\times SO(r)$-invariant,
then, up to multiplication by a constant, it coincides with
$ds^2_\kappa$ for some $\kappa$.
\end{prop}

\noindent{\it Proof.} The statement follows from the fact that the
restriction of the Hamiltonian function to $T^*_{X_0} V_{n,r}$ is
a quadratic form invariant with respect to the transformations
\eqref{SO(n-r)-action} and
$$
P_1 \longmapsto Q^{-1} P_1 Q, \quad P_2 \longmapsto P_2 Q, \quad
Q\in SO(r),
$$
where $P_1$ and $P_2$ are defined in \eqref{P-neutral}.
\hfill$\Box$

\section{The Neumann Systems on Stiefel Varieties}

The celebrated Neumann system on a sphere $S^{n-1}$ is defined as
a natural mechanical system with the quadratic Hamiltonian
(\ref{sphere_neum}). By analogy, we define {\it a Neumann on the
Stiefel variety} $V_{n,r}$ as a natural mechanical system with an
{\it $SO(n)$-invariant kinetic energy} and the quadratic potential
function
\begin{equation}
V=\frac12\tr(X^TAX)=\frac12\sum_{i=1}^r (e_i,Ae_i).
\label{Neumann_potential}
\end{equation}
Note that the above potential is constant for $r$.

While on the sphere $S^{n-1}$ an $SO(n)$-invariant kinetic energy
is unique (up to multiplication by a constant factor), on the
variety $V_{n,r}$ with $r>1$ there are many different
$SO(n)$-invariant metrics. Following Sections 3 and 4, we consider
the kinetic energy determined by the metrics $ds^2_\kappa$ (see
eq. (\ref{EM})). Thus, the Hamiltonian has the form
\begin{equation}
H_{neum,\kappa}(X,P)=\frac12\tr(P^TP)-\left(\frac12+\kappa\right)\tr((X^T P)^2)+\frac12\tr(X^TAX) \, .
\label{neumann}
\end{equation}

\begin{prop}
The Neumann system with Hamiltonian (\ref{neumann}) is given by
\begin{equation}
\begin{aligned}
& \dot X=P-(1+2\kappa) X P^T X, \\
& \dot P=-AX-XP^T P+(1+2\kappa) PX^T P+XX^TAX.
\end{aligned}\label{No}
\end{equation}
\end{prop}

\noindent{\it Proof.} It is a direct calculation. We have
\begin{equation*}
\begin{aligned}
&\dot X=\frac{\partial H}{\partial P} - X\Pi=P-(1+2\kappa) X P^T X -X \Pi, \\
&\dot P=- \frac{\partial H}{\partial X} + X\Lambda + P\Pi=-AX+(1+2\kappa) PX^T P+ X\Lambda + P\Pi.
\end{aligned}
\end{equation*}
Differentiating the constraints (\ref{cond_XP}) with respect to time gives
$$
\dot X^T X+X^T\dot X=0, \quad
\dot X^TP+X^T \dot P + \dot P^T X + P^T \dot
X = 0.
$$
The first equation implies that the Lagrange multiplier
$\Pi$ equals zero, while the second one yields $\Lambda=X^T A X-P^TP$. \hfill $\Box$
\medskip

Note that Hamiltonians (\ref{neumann}) are  right $SO(r)$-invariant, so the momentum mapping
(\ref{momentum_right_map}) is conserved by the flows (\ref{No}) for
any parameter $\kappa$. In particular, for $\kappa=0$ we get the
{\it Neumann system with the normal metric} given by
\begin{equation}
\begin{aligned}
& \dot X=P-X P^T X, \\
& \dot P=-AX+PX^T P+ X\Lambda =-AX+PX^T P-XP^T P+XX^TAX,
\end{aligned}\label{N0_XP2}
\end{equation}
while for $\kappa=-1/2$ we get the
{\it Neumann system with the Euclidean metric} with the
corresponding Hamilton equations
\begin{equation}
\begin{aligned}
&\dot X=P, \\
&\dot P=-AX+X\Lambda=-AX-XP^T P+XX^TAX.
\end{aligned}\label{NE_XP}
\end{equation}

\paragraph{The Lax pair.} Although for different $\kappa$ the flows (\ref{No}) do not coincide,
the derivatives of the momentum $\Phi$ and of the symmetric matrix $XX^T$ are the same:
\begin{equation}
\frac{d}{dt}\Phi=[XX^T,A]\, , \qquad \frac{d}{dt}(XX^T)=[\Phi,XX^T]\, .
\label{prep}
\end{equation}
As a result, the following theorem holds.

\begin{thm} \label{Neumann_LA}
Equations (\ref{No}), in particular (\ref{N0_XP2}) and
(\ref{NE_XP}),  imply the same $n\times n$ matrix Lax representation with a spectral parameter $\lambda$:
\begin{gather}
\frac{d}{dt} \mathcal L_{neum}(\lambda)=[\mathcal A_{neum}(\lambda),\mathcal L_{neum}(\lambda)]
\label{LA1} \\
\mathcal L_{neum}(\lambda)=\lambda\Phi+XX^T-\lambda^2 A, \qquad
\mathcal A_{neum}(\lambda)=\Phi-\lambda A.
\label{LA2}
\end{gather}
\end{thm}

The proof is immediate. The coefficients of the spectral curve
\begin{equation}
\Gamma\subset\mathbb{C}^2\{\lambda,\nu\}: \quad \det(\mathcal L_{neum}(\lambda)-\mu\mathbf{I}_{n})=0
\label{spectral-curve}
\end{equation}
give us the commuting integrals of both systems, which can be expressed in the form
\begin{equation}
\mathfrak F=\{ \tr(\lambda(PX^T-XP^T)+XX^T-\lambda^2 A)^k\, \vert\, k=1,\dots,n, \,
\lambda\in\mathbb{R}\}. \label{perelomov}
\end{equation}

The Lax representation (\ref{LA1}) is closely related to the
Clebsch--Perelomov rigid body system \cite{Pe}. For $r=1$ it was
given by Moser in \cite{Moser} and for $r>1$ in \cite{RS}, without
giving explicitly the equations of motion. (As was mentioned above,
the Lax pair does not define the system itself.) The book
\cite{RS} also describes the Neumann flows on Grassmannians
$G_{n,r}$, as well as the magnetic Neumann flow on $G_{n,2}$. We
shall consider these systems together with the magnetic Neumann
flows on $V_{n,2}$ in detail in Section 7 and Appendix 2,
respectively.

Alternative (or dual) Lax pairs,  which are does equivalent to
equations (\ref{NE_XP}), (\ref{N0_XP2}) (up to the action of a
finite discrete group) are given below in Section \ref{Yuri}.

\section{Compatible Poisson Brackets and Integrability}

The Neumann systems on $V_{n,r}$ admitting Lax pairs with the Lax
matrix (\ref{LA2}) can be obtained as appropriate reductions of
integrable $n$-dimensional tops having symmetric inertia tensors and
moving in a force field with a quadratic potential (the Bogoyavlenski generalization
of the Clebsch--Perelomov system \cite{Bog}), see Appendix 1 and
\cite{RS}. However, these are reductions on {\it singular}
coadjoint orbits in the corresponding Lie coalgebras, hence the
integrability of the Neumann systems does not follow directly from that of
the $n$-dimensional tops.

We now prove a non-commutative integrability of systems (\ref{No})
by using the Bolsinov condition for a set of Casimir
functions of the pencil of compatible Poisson brackets to be
complete \cite{Bo}.

\begin{thm} \label{NEUMANN} Let all the eigenvalues of $A$ be distinct. Then the Neumann systems (\ref{No}),
in particular (\ref{N0_XP2}) and (\ref{NE_XP}), are completely
integrable in the non-commutative sense with the non-commutative
set of integrals  given by (\ref{perelomov}) and by the components
of the $SO(r)$-momentum mapping (\ref{momentum_right_map}). The
generic trajectory $(X(t),P(t))$ corresponding to the maximal rank of the
momentum $\Psi$ is quasi-periodic over isotropic tori of
dimension
\begin{equation}
\frac12\left(2r(n-r)+\frac{r(r-1)}2-\left[\frac{r}{2}\right]\right)+\left[\frac{r}{2}\right].
\label{dim_fol}
\end{equation}
\end{thm}

\noindent{\it Proof.} First, we give the interpretation of the
integrals (\ref{perelomov}) from the bi-Hamiltonian point of view.
Consider the Lie algebra $gl(n)$ of $n\times n$ real matrices
equipped with the scalar product $\langle X,Y\rangle=-\frac12\tr(XY)$
and the orthogonal decomposition
${gl}(n)=so(n)+ Sym(n)$ onto skew-symmetric and symmetric matrices:
\begin{equation}
[{so}(n),{Sym}(n)]\subset Sym(n), \quad [Sym(n),Sym(n)]\subset so(n).
\label{symmetric-pair}\end{equation}
The scalar product $\langle\cdot,\cdot\rangle$ is positive definite on $so(n)$ and negative
definite on $Sym(n)$.

Let us identify $gl(n)^*$ and $gl(n)$ by means of
$\langle\cdot,\cdot\rangle$. On $gl(n)$ we
have a pair of compatible Poisson brackets given by the following Poisson tensors
\begin{eqnarray}
\begin{aligned}
&\Lambda_1(\xi+\eta,\zeta+\theta)\vert_{x}= \langle x , [\xi,\zeta]+[\xi,\theta]+[\eta,\zeta]\rangle,\\
&\Lambda_2(\xi+\eta,\zeta+\theta)\vert_x =\langle x -A,[\xi+\eta,\zeta+\theta]\rangle, \end{aligned}\label{LAMBDA12}
\end{eqnarray}
where $x \in gl(n)$, $\xi,\zeta\in so(n)$, $\eta,\theta\in Sym(n)$ (see \cite{R, Bo}).
The tensor $\Lambda_1$ corresponds to the canonical Lie--Poisson brackets in the
dual to the semi-direct product $so(n)\oplus_{\ad} Sym(n)$.

Consider the Poisson pencil
$$
\Lambda_{\lambda_1,\lambda_2}=\lambda_1 \Lambda_1+\lambda_2 \Lambda_2, \quad
\Pi=\{\Lambda_{\lambda_1,\lambda_2}
\; \vert\;  \lambda_1,\lambda_2\in \mathbb{R}, \;\lambda_1^2+\lambda_2^2\ne 0\}.
$$
For $\lambda_2\ne 0$, the bracket $\Lambda_{\lambda_1,\lambda_2}$
is isomorphic to the canonical Lie--Poisson bracket on $gl(n)$ and its Casimir functions
have the form
$$
f(x)=\tr\left(\sqrt{\frac{\lambda_2}{\lambda_1+\lambda_2}} h+ v -\frac{\lambda_2}{\lambda_1+\lambda_2} A\right)^k,
\quad k=1,\dots,n,
$$
where $x=h+v$, $h\in so(n)$, $v\in Sym(n)$ (see \cite{R, Bo}).
Let
\begin{equation}
\mathcal F=\{\tr(\lambda h+ v -\lambda^2 A)^k\, \vert\,
k=1,2,\dots,n,\, \lambda\in \mathbb{R}\}. \label{central}
\end{equation}
be the union of all the Casimir functions of the brackets
with $\lambda_1+\lambda_2\ne 0$, $\lambda_2\ne 0$. Then $\mathcal
F$ is a commutative set with respect to all the brackets in $\Pi$
 and, if the eigenvalues of $A$ are distinct,
$\mathcal F$ is a complete commutative set on a generic
symplectic leaf in $(gl(n),\Lambda_1)$ (see Theorem 1.5 in \cite{Bo}).
The mapping
$$
\Theta=\Phi+XX^T=\sum_{i=1}^r p_i \wedge e_i +\sum_{i=1}^r e_i\otimes e_i
$$
defines the Poisson mapping between $(T^*V_{n,r},
\{\cdot,\cdot\})$ and $(gl(n),\Lambda_1)$ which is invariant with
respect to the right $SO(r)$-action (\ref{right_action}) and the
transformation $(X,P)\mapsto (-X,-P)$. Indeed, $\Theta$ is the
composition of the following two Poisson mapping:
\begin{eqnarray*}
&&\Theta_1(X,P)=\Phi(X,P)+X, \quad (X,P)\in T^*V_{n,r},\\
&&\Theta_2(\xi,Y)=\xi+Y Y^T, \quad \xi\in so(n), \, Y\in M_{n,r}(\mathbb{R}).
\end{eqnarray*}
The mapping $\Theta_1$ realizes $T^*V_{n,r}$ as a coadjoint orbit
in the dual space of the semi-direct product $so(n)\oplus_\rho
M_{n,r}(\mathbb{R})$ (e.g., see equation (29.11), page 225, \cite{GS};
here $\rho$ denotes the usual multiplication of matrices) and
$\Theta_2$ is a Poisson mapping between $(so(n)\oplus_\rho
M_{n,r}(\mathbb{R}))^*$ and $(so(n)\oplus_{\ad} Sym(n))^*$ (see
Lemma 7.1 of \cite{RS}).

We have also that the algebra of integrals (\ref{perelomov}) is the pull-back of (\ref{central}):
$$
\mathfrak F=\Theta^* \mathcal F.
$$
However, the image $\Theta(T^*V_{n,r})$ is the union of {\it
singular} symplectic leaves in $(gl(n),\Lambda_1)$. Namely, the
generic symplectic leaf in $(gl(n),\Lambda_1)$ has the dimension
$n^2-n$, while the dimension of generic leaf in
$\Theta(T^*V_{n,r})$ is
\begin{equation}
2l=2r(n-r)+\frac{r(r-1)}2-\left[\frac{r}{2}\right]<n^2-n
\label{dimension}
\end{equation}
(see Lemma \ref{LEMA1} below). Nevertheless, it can be proved that
the set of the functions (\ref{central}) is complete on a generic
orbit laying in $\Theta(T^*V_{n,r})$ as well (see Lemma
\ref{LEMA2} below). That is, among the integrals $\mathcal F$ there is
at least $l$ polynomials $p_1,\dots,p_l$ independent on the
symplectic leaves in $\Theta(T^*V_{n,r})$.

The rest of the proof follows the idea of \cite{BJ2, Zu}. Namely,
since $SO(r)$ acts on $T^*V_{n,r}$ freely and preserves the
Poisson bracket $\{\cdot,\cdot\}$, the quotient space
$(T^*V_{n,r})/SO(r)$ carries natural induced Poisson bracket
$\{\cdot,\cdot\}'$. Let $\sigma: T^*V_{n,r}\to (T^*V_{n,r})/SO(r)$
be the canonical projection. Then, by definition,
\begin{equation}
\{f,g\}'(\sigma(X,P))=\{F,G\}(X,P), \quad F=f\circ\sigma, \quad G=g\circ\sigma.
\label{REDUCED_BRACKET}
\end{equation}
The Casimir functions $j_1,\dots,j_{[{r}/{2}]}$ of the brackets
$\{\cdot,\cdot\}'$ can be obtained from the $SO(r)$-invariant
functions $J_k=\tr(\Psi^{2k})$ via $j_k\circ \sigma=J_k$.

The mapping $\Theta$ induces $\mathbb{Z}_2$-Poisson covering
$$
\theta: ((T^*V_{n,r})/SO(r),\{\cdot,\cdot\}') \to
(\Theta(T^*V_{n,r}),\Lambda_1), \quad \theta\circ\sigma=\Theta.
$$
Hence the functions $p_1\circ \theta,\dots p_l\circ \theta$ are
independent on a generic symplectic leaf and, together with the
Casimir functions $j_1,\dots,j_{[{r}/{2}]}$, form a complete commutative set
of functions within $(T^*V_{n,r})/SO(r)$. In other words, the
functions
\begin{equation}
p_1\circ \Theta,\dots, p_l\circ \Theta, \quad J_1,\dots,J_{\left[\frac{r}{2}\right]}
\label{torus}
\end{equation}
form a complete commutative set of functions in the algebra of
$SO(r)$-invariant functions on $T^*V_{n,r}$. Note that the
independency of the functions $J_1,\dots,J_{[r/2]}$ at $(X,P)$ is
equivalent to the regularity of $\Ad_{SO(r)}$-orbit of
$\Psi(X,P)$. Therefore, \eqref{dim_fol} holds for the invariant
manifolds where the rank of $\Psi(X,P)$ is maximal.

Then, according to Theorem 1 in \cite{BJ4}, the functions
\eqref{torus} together with $\Psi^*(C^\infty(so(r))$
form a complete non-commutative set of functions in $T^*V_{n,r}$.
Therefore the integrals $\Theta^*\mathcal F+\Psi^*(C^\infty(so(r))$ of the Neumann system form a complete
non-commutative set. Moreover, the functions
(\ref{torus}) commute with all the integrals and therefore their
Hamiltonian flows generate generic leaves of the isotropic foliation
given by $\Theta^*\mathcal F+\Psi^*(C^\infty(so(r))$. Hence, the
dimension of the generic isotropic tori is given by (\ref{dim_fol}). \hfill$\Box$

\begin{remark}{\rm
The Hamiltonian of the {\it Neumann system with the normal metric}
has the form $H_{neum_0}=H\circ\Theta$, where
$$
H=\frac12\langle h,h\rangle -\langle A,v\rangle,
$$
defines the completely integrable (by means of the integrals (\ref{central})) Hamiltonian flow
\begin{equation}
\dot h=[v,A], \qquad \dot v=[h,v] \label{CPB}
\end{equation}
on $(so(n)\oplus_{\ad} Sym(n))^*$. This system belongs to the
class of Clebsch--Perelomov--Bogoyavlenski rigid body systems
(\cite{Pe, Bog}, see also Section \ref{grassmannian}). Note that
the representation of the Neumann system on the sphere as a system
on a adjoint orbit is given by Ratiu \cite{Ra}. }\end{remark}

\begin{lem}\label{LEMA1}
The dimension of a generic symplectic leaf in
$(\Theta(T^*V_{n,r}),\Lambda_1)$ is given by formula (\ref{dimension}).
\end{lem}

\noindent{\it Proof.} Without loss of generality, choose a
point $X_0=(E_1,\dots,E_r)\in V_{n,r}$ and a generic point $(X_0,P_0)\in
T^*_{X_0} V_{n,r}$. Denote
$$
h=\Phi(X_0,P_0), \quad v=X_0 X_0^T
$$
Then $h$ is a generic $so(n)$-matrix of the form 
$$
h=\begin{pmatrix}
h_1 & h_2
\\-h_2^T & 0
\end{pmatrix}, \quad h_1\in so(r), \quad h_2\in M_{r,n-r}(\mathbb{R}) \quad {\rm and}\quad
v=
\begin{pmatrix}
\mathbf{I}_r & 0 \\
0 & 0
\end{pmatrix}.
$$

>From the definition (\ref{LAMBDA12}) of the tensors $\Lambda_1,
\Lambda_2$ we get $\xi+\eta \in \ker \Lambda_1(h+v)$, $\xi\in
so(n)$, $\eta\in Sym(n)$ if and only if
\begin{eqnarray}
&&[\xi,v]=0, \label{L1}\\
&&[\xi,h]+[\eta,v]=0. \label{L2}
\end{eqnarray}
The first equation gives the condition for $\xi$ to belongs to the subalgebra $so(r)\oplus so(n-r)$.
Denote
$$
\xi=
\begin{pmatrix}
\xi_1 & 0 \\
0 & \xi_2
\end{pmatrix}, \quad
\eta=
\begin{pmatrix}
\eta_1 & \eta_3 \\
\eta_3^T & \eta_2
\end{pmatrix}, \quad
$$
where $\xi_1\in so(r)$, $\xi_2\in so(n-r)$, $\eta_1\in Sym(r)$, $\eta_2\in Sym(n-r)$, $\eta_3\in M_{r,n-r}(\mathbb{R})$.
Since
$$
[\eta,v]=
\begin{pmatrix}
0 & -\eta_3 \\
\eta_3^T & 0
\end{pmatrix},
$$
from (\ref{symmetric-pair}) and (\ref{L2}) we find that
$[\xi_1,h_1]=0,$, $\xi_2$, $\eta_1$, $\eta_2$ are arbitrary, and $\eta_3$ is uniquely determined from the
equation
$$
\left[\begin{pmatrix}
\xi_1 & 0 \\
0 & \xi_2
\end{pmatrix},
\begin{pmatrix}
0 & h_2 \\
-h_2^T & 0
\end{pmatrix}\right]
+\begin{pmatrix}
0 & -\eta_3 \\
\eta_3^T & 0\end{pmatrix}=0.
$$
For generic $h_1$, the solutions of $[h_1,\xi_1]=0$ form a maximal
commutative subalgebra of $so(r)$. That is, the dimension of the
space of the solution of (\ref{L1}), (\ref{L2}) is
$$
\dim\ker\Lambda_1(h+v)=\rank so(r) +\dim so(n-r) + \dim Sym(r)+\dim Sym (n-r).
$$
Finally, the dimension of a generic symplectic leaf in
$(\Theta(T^*V_{n,r}),\Lambda_1)$ is equal to
$n^2-\dim\ker\Lambda_1(h+v)=2l$. \hfill$\Box$

\begin{lem}\label{LEMA2}  If all eigenvalues of $A$ are distinct, then
the set of the functions (\ref{central}) is a complete commutative set
on a generic symplectic leaf in $(\Theta(T^*V_{n,r}),\Lambda_1)$
\end{lem}

\noindent{\it Proof.} We keep the notation from the proof of Lemma \ref{LEMA1}.
The proof presented here is a modification of that of Theorem 1.6 in \cite{Bo}.
According to Theorem 1.1 in \cite{Bo}, the set of the functions
(\ref{central}) is complete on the symplectic leaf containing the point $x=h+v$ if and only if

\begin{description}
\item{(A1)} All the brackets $\Lambda_{\lambda_1,\lambda_2}$ non-proportional to $\Lambda_1$
have the maximal rank $n^2-n$ at $x$.

\item{(A2)} The kernel of the bracket $\Lambda_{1,-1}$ at $x$, restricted to the linear space
$\ker\Lambda_{1}$, has dimension $n$,
\begin{equation}
\dim\{(\xi+\eta\in \ker\Lambda_1(x)\, \vert\, \Lambda_{1,-1}(\xi+\eta,\ker\Lambda_{1}(x))\vert_x=0\}.
\label{A2}
\end{equation}
\end{description}

Here all the objects are assumed to be complexified. Since the
conditions (A1) and (A2) are both generic, it is sufficient to find
$x_1\in \Theta(T^*V_{n,r})$ for which (A1) holds and $x_2\in \Theta(T^*V_{n,r})$ which satisfies (A2).
Then the set of $x\in \Theta(T^*V_{n,r})$ satisfying both conditions
will be open and dense everywhere in the induced topology on $\Theta(T^*V_{n,r})$.

If $\lambda_1+\lambda_2\ne 0$, $\lambda_2\ne 0$, then the bracket $\Lambda_{\lambda_1,\lambda_2}$
is isomorphic to the canonical Lie--Poisson bracket on $gl(n,\mathbb{C})$. Thus the
brackets $\Lambda_{\lambda_1,\lambda_2}$ ($\lambda_1+\lambda_2\ne 0$, $\lambda_2\ne 0$)
have the maximal rank $n^2-n$ at $x=h+v$ if and only if the complex line
$$
\mathcal L=\{h+v-\lambda A\, \vert\, \lambda\in\mathbb{C}\}
$$
intersects the set of singular points of $gl(n,\mathbb{C})$ only at $x=h+v$.
This condition is obviously satisfied if all the eigenvalues of $A$ are distinct.

To prove (A1) we have to find the (complex) dimension of
$\ker\Lambda_{1,-1}$. From the definition (\ref{LAMBDA12}) we get
\begin{equation}
\Lambda_{1,-1}(\xi+\eta,\zeta+\theta)\vert_{h+v} =
-\langle h,[\eta,\theta]\rangle+\langle A,[\xi,\theta]+[\eta,\zeta]\rangle
\label{LAMBDA}
\end{equation}
and $\xi+\eta \in \ker \Lambda_{1,-1}(h+v)$, $\xi\in so(n,\mathbb{C})$, $\eta\in Sym(n,\mathbb{C})$
if and only if
\begin{eqnarray}
&&[\eta,A]=0, \label{L3}\\
&&[\xi,A]-[\eta,h]=0. \label{L4}
\end{eqnarray}
The solutions of (\ref{L3}) are all diagonal matrices. For the given diagonal matrix $\eta$,
the matrix $\xi$ is uniquely determined from (\ref{L4}). Therefore, $\dim\ker\Lambda_{1,-1}(h+v)$.
\medskip

It remains to check the condition (A2). Take $h$ of the form
$$
h=\begin{pmatrix}
h_1 & 0
\\0 & 0
\end{pmatrix},
$$
$h_1$ being a generic element of $so(r)$. From the proof of Lemma \ref{LEMA1} we have
\begin{equation}
\ker \Lambda_{1}(h+v)=so(r,\mathbb{C})_{h_1}\oplus Sym(r,\mathbb{C})\oplus gl(n-r,\mathbb{C})\, ,
\label{KER}
\end{equation}
where $so(r,\mathbb{C})_{h_1}=\{\xi_1\in so(r,\mathbb{C})\, \vert\, [\xi_1,h_1]=0\}$ is a Cartan subalgebra
of $so(r,\mathbb{C})$.

>From (\ref{LAMBDA}), (\ref{KER}) we conclude that $\xi+\eta$
belongs to $\ker\Lambda_{1,-1}(x)\vert_{\ker\Lambda_{1}(x)}$ if
and only if
\begin{eqnarray}
&&[A_1,\xi_1]-[h_1,\eta_1]=0,\label{L5}\\
&&[A_1,\eta_1] \in so(r,\mathbb{C})^\perp,\label{L6} \\
&& \xi_2=0,\label{L7}\\
&&[A_2,\eta_2]=0\label{L8},
\end{eqnarray}
where $\eta_1$ and $\eta_2$ are defined in the proof of Lemma \ref{LEMA1} and
$$
A_1=\diag(a_1,\dots,a_r), \quad A_2=\diag(a_{r+1},\dots,a_n).
$$

Equations (\ref{L5}), (\ref{L6}) form a closed system within
$gl(r,\mathbb{C})$, and from the proof of Theorem 1.6 \cite{Bo},
the dimension of solution of (\ref{L5}), (\ref{L6}) is equal to
$r$. On the other side, the solution of (\ref{L8}) are all
diagonal matrices in $gl(n-r,\mathbb{C})$. Hence (\ref{A2}) holds.
\hfill$\Box$

\paragraph{Singular matrices $A$ and the Einstein metrics.}
Now suppose that not all the eigenvalues of $A$ are distinct:
$$
a_1=\dots=a_{k_1}, \,\,a_{k_1+1}=\dots=a_{k_1+k_2},\,\, \dots, \,\, a_{n+1-k_r}=\dots=a_n,
\quad k_1+k_2+\dots+k_r.
$$
Then we have a non-trivial isotropy algebra
$$
so(n)_A=\{\xi\in so(n) \, \vert\, [\xi,A]=0\}=so(k_1)\oplus so(k_2)\oplus\dots\oplus so(k_r).
$$
Let $\mathcal G$ be the set of linear functions on $so(n)_A$. The
set $\mathcal F+\mathcal G$ is a complete non-commutative set of
function on $(gl(n),\Lambda_1)$ (Theorem 1.5 \cite{Bo}). By
modifying the proof of Lemma \ref{LEMA1} and Bolsinov's Theorem
1.5 \cite{Bo}, one can prove that the set of functions $\mathcal
F+\mathcal G$ is complete on $\Theta(T^*V_{n,r})$ as well,
implying non-commutative integrability of the Neumann systems
(\ref{No}). The complete verification is out the scope of this
paper.

Let $SO(n)_A=SO(k_1)\times SO(k_2)\times \dots\times SO(k_r)\subset SO(n)$ be the
adjoint isotropy group of $A$. The momentum mapping of
the left $SO(n)_A$-action is given by
$$
\Phi_A=\pr_{so(n)_A}\Phi=\pr_{so(n)_A}\left( PX^T-XP^T\right)
$$
and $\Theta^*\mathcal G$ are exactly Noether integrals arising from the $SO(n)_A$-symmetry
of the Neumann flows.

In particular, when
$A=0$, we get the integrals of the geodesic flow of the metric $ds^2_\kappa$ in the form
$\Phi^*(C^\infty(so(n))+\Psi^*(C^\infty(so(r))+\mathfrak F$, where now
\begin{equation} \mathfrak F=\{ \tr(\lambda(PX^T-XP^T)+XX^T)^k\, \vert\,
k=1,\dots,n, \, \lambda\in\mathbb{R}\}.\label{FF} \end{equation}

We shall mention the following important corollary of the above
construction.

Let $\mathbb B: so(r) \to so(r)$ be positive definite and $h_\mathbb B=\frac12\langle \xi,\mathbb B\xi\rangle$.
Suppose that the Euler equations
\begin{equation}
\dot f=\{f,h_\mathbb B\}^-_{so(r)} \quad \Longleftrightarrow \quad
\dot \xi=[\xi, \nabla h_\mathbb B(\xi)]=[\xi, \mathbb B \xi], \quad \xi\in so(r)
\label{e-so(r)}
\end{equation}
are completely integrable with a complete commutative set of functions $\mathcal B$.

\begin{thm}\label{CIF}
\begin{description}
\item{\rm{(i)}} $\mathfrak B+\mathfrak F$ is a complete
commutative set of $SO(n)$-invariant functions on $T^*V_{n,r}$,
where $\mathfrak F$ is given by (\ref{FF}) and $\mathfrak
B=\Psi^*(\mathcal B)$.

\item{\rm{(ii)}} The geodesic flow of the $SO(n)$-invariant metric
$ds^2_{\mathbb B}$ on $V_{n,r}$ defined by the Hamiltonian
function
\begin{equation*}
H_{\mathbb B}=\frac12\tr(P^T P)-\frac{1}{4}\tr\left((X^TP-P^TX)\mathbb B(X^TP-P^TX)\right)
\label{HB}\end{equation*}
is completely integrable in the non-commutative
sense. The complete set of first integrals is
$
\Phi^*(C^\infty(so(n))+\mathfrak F+\mathfrak B ,
$
and the generic trajectories of the system are quasi-periodic over the isotropic tori
of dimension
$$
\ddim(\mathfrak F+\mathfrak B)=\dim V_{n,r}-l.
$$
Here $2l$ is the dimension of a generic adjoint orbit in
$\Phi(T^*V_{n,r})$ (see Lemma \ref{2l}).
\end{description}
\end{thm}

For example, if the matrix $B=\diag(B_1,\dots,B_r)$ has distinct eigenvalues, we can take
$\mathbb B(\xi)=B\xi+\xi B$ (the Manakov operator on $so(r))$ and the  commutative set
\begin{equation}
\mathfrak B=\{\,\tr(\Psi+\mu B)^k, \, \mu\in \R, \, k=1,2,\dots,r\,\}.
\label{BB}
\end{equation}

Now, by using the chain method for the construction of complete commuting sets of functions on Lie algebras developed by Mikityuk \cite{Mik}, one can easily prove that the Hamiltonian function (\ref{EM3}) defines
completely integrable system on $so(r)$. Therefore we get

\begin{cor}\label{nikinorov}
The $SO(n)$-invariant geodesic flows determined by Hamiltonian functions (\ref{EM2})
are completely integrable. In particular, the geodesic flow of the
Einstein metrics constructed in \cite{ADN} are completely integrable.
\end{cor}

Note that the set of functions given in Theorem \ref{CIF} differs
from those described in Theorem 2.6. Another proof of integrability of geodesic flows of the
Einstein metrics, based on using singular Manakov flows, is recently obtained in \cite{DGJ}.

\paragraph{Commutative integrability of the Neumann flows.}
We turn back to the Neumann flows. According to Theorem
\ref{NEUMANN}, systems (\ref{No}) are integrable in the
noncommutative sense by means of the integrals
$\Psi^*(C^\infty(so(r))+\mathfrak F$, where $\mathfrak F$ is given
by (\ref{perelomov}). However, the Neumann flows ({\ref{No}) are
integrable in the commutative (Liouville) sense as well: instead of
$\Psi^*(C^\infty(so(r)))$  one should take, for example,
the commutative set (\ref{BB}). Moreover, it follows:

\begin{cor}\label{COM}
Let all the eigenvalues of $A$ be distinct.
Suppose that the Euler equations (\ref{e-so(r)}) are completely integrable with a
complete commutative set of functions $\mathcal B$.
Then the Neumann system with the kinetic energy given by the $SO(n)$-invariant metric $ds^2_{\mathbb B}$
$$
H_{neum,\mathbb B}=\frac12\tr(P^T P)-\frac{1}{4}\tr\left((X^TP-P^TX)\mathbb B(X^TP-P^TX)\right)+\frac12\tr(X^TAX)
$$
is completely integrable.
The complete commutative set of first integrals is
$\mathfrak F+\mathfrak B$,
where $\mathfrak F$ is given by (\ref{perelomov}) and $\mathfrak B=\Psi^*(\mathcal B)$.

In particular, the Neumann systems with the kinetic energy determined by $SO(n)$-invariant Einstein metrics
constructed in \cite{ADN} are completely integrable.
\end{cor}

\section{Reduction to Grassmannians}
\label{grassmannian}

By definition, the points of the oriented Grassmannian variety
$G_{n,r}$ are $r$-dimensional oriented planes passing through the
origin in the Euclidean space $\R^n$. The usual action of the
group $SO(n)$ on $\R^n$ yields a transitive action on the set of
all $r$-dimensional planes, i.e., on $G_{n,r}$. The isotropy group
of the $r$-plane spanned by the vectors $E_1, \dots, E_r$ (relative to
the base (\ref{base})) has the form
$$
\begin{pmatrix}
SO(r) & 0 \\
0 & SO(n-r)
\end{pmatrix}\cong SO(r)\times SO(n-r).
$$
It follows that $G_{n,r}\cong SO(n)/(SO(r)\times SO(n-r))$.

The oriented Grassmannian can also be seen as a quotient space of
the Stiefel manifold by the right $SO(r)$-action described in
Section 1. The quotient mapping $V_{n,r}\to G_{n,r}$ is
$$
X=(e_1,\dots,e_r) \longmapsto e_1 \wedge \dots \wedge e_r.
$$

The symplectic leaves in $(T^*V_{n,r})/SO(r),\{\cdot,\cdot\}')$
with $\{\cdot,\cdot\}'$ given by (\ref{REDUCED_BRACKET}) are the
Marsden--Weinstein symplectic reduced spaces of $T^* V_{n,r}$. In
particular, the reduced space that corresponds to zero value of
the momentum mapping
$$
\Psi^{-1}(0)/SO(r), $$ is symplectomorphic to the cotangent bundle
$T^* G_{n,r}$ equipped with a canonical symplectic structure. Note
that $(X,P)$ belongs to $\Psi^{-1}(0)$ if and only if $X^T P=0$.

The last condition also implies that, although the Neumann systems
(\ref{N0_XP2}) and (\ref{NE_XP}) are different on the whole
$T^*V_{n,r}$, they coincide on $\Psi^{-1}(0)$, hence their
reductions onto the cotangent bundle $T^*G_{n,r}$ are the same.

The reduced flow can be written in the alternative Euler--Lagrange
form. Namely, we have
$$
\frac{d}{dt}\left(e_1 \wedge \dots \wedge e_r\right)= \sum_{i=1}^r e_1 \wedge \dots \wedge \dot e_i \wedge \dots \wedge e_r
$$
and, in view of the matrix equation (\ref{NE_XP}), the reduced system is
\begin{eqnarray}
&& \frac{d^2}{dt^2}\left(e_1 \wedge \dots \wedge e_r\right)=-\sum_{i=1}^r e_1 \wedge \dots \wedge Ae_i \wedge \dots \wedge e_r \nonumber\\
&&\qquad +2\sum_{1\le i<j\le r} e_1\wedge \dots\wedge \dot e_i \wedge \dots \wedge \dot e_j \wedge \dots \wedge e_r +\lambda \left(e_1 \wedge \dots \wedge e_r\right), \label{neum-gras}\\
&&\qquad \lambda=\tr\Lambda=\tr(X^TAX-\dot X^T \dot X)=\sum_{i=1}^r \left( ( e_i,Ae_i)-( \dot e_i,\dot e_i ) , \right).\nonumber
\end{eqnarray}
where
\begin{equation}
\label{gras-velocities} (e_i,e_j)=\delta_{ij}, \quad  (\dot e_i,e_j)=0, \quad  i,j=1,\dots,r.
\end{equation}

Note that (\ref{neum-gras}) is $SO(r)$-invariant. We refer to
(\ref{neum-gras}) as a {\it Neumann system} on the oriented
Grassmannian variety $G_{n,r}$.

\begin{thm}\label{NEUMANN2}
Suppose that all the eigenvalues of $A$ are distinct. Then the Neumann
system on $T^* G_{n,r}$ is completely integrable in the Liouville
sense by means of the integrals induced from the $SO(r)$-invariant
functions (\ref{perelomov}).
\end{thm}

\noindent{\it Proof.} Since $T^*G_{n,r}=\Psi^{-1}(0)/SO(r)$ is a
singular symplectic leaf in $T^*V_{n,r}/SO(r)$,  the statement of
the theorem does not follow directly from Theorem \ref{NEUMANN}.

We keep the notation from the proofs of Theorem \ref{NEUMANN},
Lemma \ref{LEMA1} and Lemma \ref{LEMA2}. Let $h=\Psi(X_0,P_0)$,
$v=X_0X_0^T$, where $(X_0,P_0)\in \Psi^{-1}(0) \cap T_{X_0}^*
V_{n,r}$ is in a generic position. Then $h_1=0$. From the proof of
Lemma \ref{LEMA1} we get
$$\dim\ker\Lambda_1(h+v)=\dim (gl(r)\oplus gl(n-r)).$$
and the mappings $\Theta$ and $\theta$ map $\Psi^{-1}(0)$ and $\Psi^{-1}(0)/SO(r)$
to the single symplectic leaf in $(gl(n),\Lambda_1)$.
Now, by modifying the proof of Lemma \ref{LEMA1} one can prove conditions (A1) and (A2) for $x=h+v$ with $h_1=0$ as well.
Therefore, among functions (\ref{central}) one can find exactly
$$
r(n-r)=\frac12(\dim gl(n)-\dim gl(r)-\dim gl(n-r))=\dim G_{n,r}
$$
independent functions $p_1,\dots,p_{r(n-r)}$.
Thus the functions $p_1\circ\theta,\dots,p_{r(n-r)}\circ\theta$ provide a complete commutative
set of functions on $\Psi^{-1}(0)/SO(r)$. \hfill$\Box$

\paragraph{The special case $r-1$.}
The Stiefel variety $V_{n,n-1}$ is diffeomorphic to
$SO(n)=V_{n,n}$: to $e_1,\dots,e_{n-1}$ one can associate the
unique unit vector $e_n$ such that $e_1,\dots,e_n$ is the
orthonormal base with the same orientation as $E_1,\dots,E_n$:
\begin{equation}
X=(e_1,\dots,e_{n-1}) \longmapsto \mathcal X=(e_1,\dots,e_{n-1},e_n)\in SO(n).
\label{iden1}
\end{equation}
Similarly, the oriented Grassmannian variety $G_{n,n-1}$ is
diffeomorphic to $G_{n,1}=S^{n-1}$ via the mapping
\begin{equation}
e_1\wedge \dots \wedge e_{n-1} \longmapsto  e_n.\label{iden2}
\end{equation}

It is natural to expect that in this case the Neumann system
(\ref{neum-gras}) gives rise to the classical Neumann system on
the sphere $S^{n-1}$. Indeed, due to conditions
(\ref{gras-velocities}), for $r-1$ the second term in the
right-hand side of equations (\ref{neum-gras}) vanishes. Then,
under identification (\ref{iden2}), these equations gives rise to
\begin{equation}
\ddot e_n=(A- \tr A {\bf I}) e_n + \lambda e_n=A e_n-\left( ( e_n,A e_n )+( \dot e_n,\dot e_n ) \right) e_n \,,
\label{neum-gras**}
\end{equation}
which describes the motion on the sphere $S^{n-1}=\{\langle e_n,e_n\rangle=1 \}$
with the potential $-\frac12 \langle A e_n, e_n \rangle$.

\paragraph{The 4-th degree potential.}
By using Cartan models of symmetric spaces, a class of new
integrable potential systems on such spaces was obtained by
Saksida \cite{Sa}. As noticed in \cite{Sa2}, in the case of the
sphere $S^{n-1}$, such a system is a generalization of the Neumann
system in presence of a potential of degree 4. The latter system
is separable in the spherical elliptic coordinates and was found
previously in \cite{Wo}.

In addition, it can be proved that the construction of \cite{Sa}
on the Grassmannian varieties gives rise to the Hamiltonian
\begin{equation} \label{quartic}
H(X,P)=\frac12 \tr(P^TP)+\tr (X^T A^2 X)-\tr(X^TAXX^TAX) ,
\end{equation}
which for $r=1$, takes the well known form
$H=\frac12(p,p)+\sum_{i=1}^n a_i^2 e_i^2-\left(\sum_{i=1}^n a_i
e_i^2 \right)^2$ \cite{Wo, Sa2}. For $r>1$, the Hamiltonian flow
on $T^*V_{n,r}$ determined by the Hamiltonian function
(\ref{quartic}) is integrable after its restriction to the
invariant manifold $\Psi^{-1}(0)\subset T^*V_{n,r}$ and the
reduction to $T^*G_{n,r}$. This system will be discussed
elsewhere.

\section{The Dual Lax Pair and Geometric Interpretation of Integrals} \label{Yuri}

As mentioned in Section 5, like the classical Neumann system on the sphere $S^{n-1}$,
the Neumann systems on $V(r,n)$ also admit dual Lax representations.

\begin{thm} \label{DUAL}
Up to the action of a discrete group $\mathbb Z_2^n$ generated by
reflections (\ref{reflections}), the
Neumann flows (\ref{N0_XP2}) and (\ref{NE_XP}) are equivalent to
the following $2r\times 2r$ matrix Lax pair with a rational
spectral parameter $\lambda$
\begin{gather}
\frac{d}{dt} \mathcal L^*_{neum}(\lambda)
=[\mathcal L^*_{neum}(\lambda),\mathcal A^*_{neum}(\lambda)], \label{LA3} \\
\mathcal L^*_{neum}(\lambda)=\begin{pmatrix}
-X^T( \lambda \mathbf{I}_n-A)^{-1} P & -X^T(\lambda \mathbf{I}_n-A)^{-1} X \\
\mathbf I_r+P^T(\lambda \mathbf{I}_n- A)^{-1} P & P^T(\lambda \mathbf{I}_n-A)^{-1} X
\end{pmatrix} ,
\end{gather}
where for system (\ref{N0_XP2}), respectively (\ref{NE_XP}), one
should put
\begin{equation}
\mathcal A^*_{neum}(\lambda)=\begin{pmatrix}
X^TP & \mathbf{I}_r \\
\Lambda-\lambda \mathbf{I}_r & - P^T X
\end{pmatrix}, \quad \textup{respectively,  }  \quad
\mathcal A^*_{neum}(\lambda)=\begin{pmatrix}
0 & \mathbf{I}_r \\
\Lambda-\lambda \mathbf{I}_r & 0
\end{pmatrix},
\label{LA4}
\end{equation}
where $\Lambda=X^T A X-P^TP$.
\end{thm}

The statement is checked straightforwardly by using constraints
(\ref{cond_XP}) and the matrix identities
$$
A(\lambda \mathbf{I}_n-A)^{-1}= (\lambda \mathbf{I}_n-A)^{-1}A=\lambda(\lambda \mathbf{I}_n-A)^{-1} -\mathbf{I}_n\, .
$$

The dual Lax pair (\ref{LA4}) for the generalized Neumann system
(\ref{NE_XP}) was first given in unpublished manuscript
\cite{Kap89}. For $r=1$ it gives the known $2\times 2$ Lax pair
for the Neumann system indicated in several publications (see,
e.g., \cite{Su} and references therein).

\begin{remark}{\rm
For $r$ the considered Neumann flows become geodesic flows of
bi-invariant metrics on $SO(n)$. Then both Lax representations
($n\times n$ and $2n\times 2n$) give the integrals for the
geodesic flow of a bi-invariant metric, but also they give the
integrals for the right-invariant Manakov geodesic flows on
$SO(n)$ described by equations (\ref{dot_XP2}), (\ref{Manakov}).
}\end{remark}

\paragraph{The spectral curve and integrals.} Let
$$
{a}(\lambda) =(\lambda-a_1)\cdots (\lambda-a_n).
$$

The spectral curve of $\mathcal L^*_{neum}(\lambda)$ can be written in form
$$
|a(\lambda) \mathcal L^*_{neum}(\lambda)-w{\bf I}_n|\equiv
w^{2r} + w^{2r-2} {a}(\lambda){\cal I}_{2}(\lambda)+ \cdots +w^2 a^{2r-3}(\lambda) {\cal I}_{2r-2}(\lambda)
+ a^{2r-1} {\cal I}_{2r}(\lambda)=0 ,
$$
where ${\cal I}_{2l}(\lambda)$ are invariant polynomials in the components of the wedge products \\
$e_{j_1}\wedge \cdots \wedge e_{j_j}$ and
\begin{align}
e_1 & \wedge \cdots \wedge e_r, \nonumber \\
e_1 & \wedge \cdots \wedge e_r \wedge p_i, \quad i=1,\dots, r, \nonumber \\
\cdots & \cdots \cdots \cdots \label {forms}  \\
e_1 & \wedge \cdots \wedge e_r \wedge p_1\wedge\cdots\wedge p_r . \nonumber
\end{align}
Note that, due to the symplectic block structure
of $\mathcal L^*_{neum}(\lambda)$, the coefficients at odd powers of $w$ in the spectral curve are zero.

In the case $2r\le n$ the polynomials can be written in form
\begin{align}
{\cal I}_{2}(\lambda) & = \sum_{i=1}^n \frac{ a(\lambda) }{\lambda-a_i } \left( (e^i_1)^2 + \cdots+ (e^i_r)^2\right)
 + \sum_{1\le i<j\le n} \frac{a(\lambda) }{(\lambda-a_i)(\lambda-a_j)} \, \Phi_{ij}^2 \, , \nonumber \\
& \cdots \quad \cdots \nonumber \\
{\cal I}_{2l}(\lambda) & = \sum_{I_l} \frac{{a}(\lambda)}{(\lambda-a_{i_1} )\cdots(\lambda-a_{i_l})}
\sum_{J_l=\{j_1,\dots,j_l \}} ( e_{j_1}\wedge \cdots \wedge e_{j_l} )_{I_l}^2 \nonumber \\
& \quad + \sum_{I_{l+1}} \frac{{a}(\lambda)}{(\lambda-a_{i_1}) \cdots (\lambda-a_{i_{l+1}})}
\sum_{J_l, 1\le i \le r } (e_{j_1} \wedge \cdots \wedge e_{j_l}\wedge p_i )_{I_{r+1}}^2  + \cdots , \nonumber  \\
& \cdots \quad \cdots \nonumber \\
{\cal I}_{2r}(\lambda)
& = \sum_{I_r} \frac{{a}(\lambda)}{(\lambda-a_{i_1})\cdots (\lambda-a_{i_r})} ( e_1\wedge \cdots \wedge e_r )_{I_r}^2  \nonumber \\
& \quad +\sum_{I_{r+1} } \frac{{a}(\lambda)}{(\lambda-a_{i_1}) \cdots (\lambda-a_{i_{r+1}}  )  }
\sum_{i=1}^r (e_1 \wedge \cdots \wedge e_r\wedge p_i)_{I_{r+1}}^2 \nonumber \\
& \quad + \sum_{I_{r+2}}
\frac{ {a}(\lambda) }{(\lambda-a_{i_1})
\cdots (\lambda-a_{i_{r+2}})} \sum_{1\le i<j\le r} (e_1\wedge \cdots \wedge e_r\wedge p_i\wedge p_j)_{I_{r+2}}^2 + \cdots \nonumber \\
& \quad +\sum_{I_{2r}} \frac{ {a}(\lambda) }{ (\lambda-a_{i_1})\cdots (\lambda-a_{i_{2r}} ) } \,|\Phi|^{I_{2r}}_{I_{2r}},
\label{2r-integrals}
\end{align}
where $I_k=\{i_1,\ldots i_k\}\subset\{1,\dots,n\}$ is the
multi-index with distinct indices $1\le i_1<\cdots<i_k\le n$ and $
|\Phi|^{I_k}_{I_k}$ is the $k\times k$ diagonal minor of the
momentum matrix $\Phi$ corresponding to the multi-index $I_k$.
Note that, in view of definition of $\Phi$,
$$
|\Phi|^{I_{2r}}_{I_{2r}} = (e_1\wedge \cdots \wedge e_r\wedge p_1\wedge \cdots \wedge p_r)_{I_{2r}}^2 .
$$

In the case $2r>n$ the polynomials ${\cal I}_{2l}(\lambda)$ have the same form with the only difference:
the terms with the wedge products of $e_i, p_j$ of order $>n$ are absent.

It follows that in both cases ${\cal I}_{2l}(\lambda)$ are polynomials in $\lambda$ of degree $n-l$ and that
the leading coefficients of ${\cal I}_{2}(\lambda), \dots, {\cal I}_{2r}(\lambda)$
produce trivial constants on $V(r,n)$. Hence, as a simple
counting shows, the  number of nontrivial integrals on $T^*V(r,n)$ provided by the Lax matrix
$\mathcal L^*_{neum}(\lambda)$ in (\ref{LA4}) equals
$$ N=(n-1)+(n-2)+\cdots+(n-r)=r(n-r)+ r(r-1)/2, $$
which coincides with the dimension of the Stiefel variety.

Note that, although the Lax matrix $\mathcal L^*_{neum}(\lambda)$ is not invariant under the right $SO(r)$-action,
the spectral curve and therefore all the integrals $\mathcal I_{2l}(\lambda)$ are $SO(r)$-invariant.

Since the number $N$ is bigger than half of dimension
(\ref{dimension}) of a generic symplectic leaf within
$(T^*V_{n,r})/SO(r)$, some of the integrals are dependent. Like
the "big" Lax matrix ${\mathcal L}_{neum}(\lambda)$ in
(\ref{LA1}), the dual Lax matrix $\mathcal L^*_{neum}(\lambda)$
does not produce explicitly the momenta integrals $\Psi_{ij}$. 

\paragraph{Geometric interpretation of the integrals ${\cal I}_{2r}(\lambda)$.}
The components of forms (\ref{forms}) that appear in the last
invariant polynomial ${\cal I}_{2r}(\lambda)$ have a transparent
geometric interpretation: they are Pl\"ucker coordinates of the
$2r$-dimensional linear subspace ($2r$-plane)
$$
\bar\Sigma=\bar\Sigma (X,P) \subset {\mathbb
R}^{n+r}(x_1,\dots,x_n,y_1,\dots,y_r)
$$
spanned by the columns of the $2r\times (n+r)$ matrix
\begin{equation}
{\mathcal V} = \mathcal V(X,P)=\begin{pmatrix}  e_1 & \cdots & e_r & p_1 & \cdots & p_r \\
  0 & \cdots & 0 &   & & \\
\vdots &  & \vdots & & {\mathbf{I}_{r}} & \\
0   &  \cdots & 0 & &&
\end{pmatrix} ,
\label{VV}
\end{equation}
${\mathbf{I}_{r}}$ being the identity
$r\times r$ matrix. Indeed (see, e.g., \cite{Griffits}), for any
$k,m \, (k<m)$,
 the Pl\"ucker coordinates of a $k$-plane $\pi$ in ${\mathbb R}^m(x_1,\dots,x_m)$
spanned by  independent vectors $v_1,\dots,v_k \in {\mathbb R}^m$ are the coefficients $G_I$ of the polynomial
$$
v_1\wedge\cdots \wedge v_k = \sum_I G_I \, dx_{i_1}\wedge\cdots \wedge dx_{i_k} ,
$$ where $I= \{i_1,\ldots i_k\} \subset\{1,\dots,n\}$ is the multi-index with $1\le i_1<\cdots<i_k\le n$.

Then the Pl\"ucker coordinates of $\bar\Sigma$ are given by all $2r\times 2r$ minors
of $\mathcal V$. In particular, the $2r\times 2r$ minors that completely contan $\mathbf I_{r}$ give
the Pl\"ucker coordinates of the $r$-plane span($e_1,\dots, e_r)\subset {\mathbb R}^{n}$.

Now consider the following family of confocal cones in ${\mathbb R}^{n+r}(x_1,\dots,x_n,y_1\dots, y_r)$
\begin{equation} \label{conf0}
\bar Q(\lambda) = \left\{ \frac{x_1^2}{\lambda - a_1}+\cdots+ \frac{x_n^2}{\lambda- a_n}+ y_1^2 +\cdots + y_r^2 =0
\right \}, \quad \lambda\in {\mathbb R}.
\end{equation}

The following theorem is a first variant of a generalization of the remarkable Chasles theorem describing a
geometric relation between the geodesic
flow on an ellipsoid and common tangent lines of confocal quadrics (\cite{Chasles, Moser, Knorr1}).

\begin{thm} \label{Gen_Chasles}
Let the $2r$-plane $\bar\Sigma(t)\subset {\mathbb R}^{n+r}$ be
associated to a generic solution $(X(t), P(t))$ of the Neumann
systems (\ref{N0_XP2}) and (\ref{NE_XP}) on $V_{n,r}$ as described
above. Then $\bar\Sigma(t)$ is tangent simultaneously to $n-r$
fixed confocal cones $\bar Q(c_1), \dots, \bar Q(c_{n-r})$, where
$c_1,\dots, c_{n-r}$ are the roots of the invariant polynomial
${\cal I}_{2r}(\lambda)$.
\end{thm}

One can show that for real solutions $(X(t), P(t))$ all these
cones are real.

In the particular case $r=1$, one can also consider the section of
$\bar\Sigma$ and of the family $\bar Q(\lambda)$ by the subspace
$\{y_1=1\}\cong {\mathbb R}^{n}$, which give respectively an
affine line $l(t)=p(t)+ \Span \{ e(t) \}$ and the family of
confocal quadrics
$$
Q(\lambda)=  \left\{ \frac{x_1^2}{a_1-\lambda}+\cdots+ \frac{x_n^2}{a_n-\lambda}=1\right\} .
$$
Then, due the above theorem, $l(t)$ is tangent to $n-1$ fixed
quadrics $Q(c_1), \dots, Q(c_{n-r})$, and we recover the following
variant of the Chasles theorem given for the Neumann
system\footnote{It can be formulated in two different ways (see
Theorem 12 in \cite{Fe} given for the Clebsch--Perelomov systems
and Theorem 4.10 in \cite{Mum}).}

\begin{prop} \textup{(Moser, \cite{Moser})} \label{Mum_geod}
\begin{description}
\item{\rm (i)}
Let $(e(t), p(t))$ be a solution of the system on $T^*S^{n-1}$ with the Hamiltonian
$$
{\mathcal H}= \sum_{i=1}^n \alpha_i F_i, \qquad F_i= e_i^2 + \sum_{j\ne i} \frac {(e_i p_j-e_j p_i)^2 }{a_i-a_j},
$$
$\alpha_i$ being arbitrary constants.
Then the associated line $l(t)$ is tangent to $n-1$ fixed confocal
quadrics of the family $Q(\lambda)$.

\item{\rm (ii)}
If  $(e(t), p(t))$ is a solution  of the system on $T^*S^{n-1}$ with ${\mathcal H}= \sum_{i=1}^n F_i/a_i$ restricted to ${\mathcal H}=0$,
the corresponding line $l(t)$ is tangent to the ellipsoid $Q(0)$, on which the contact point
$l\cap Q(0)$ traces a geodesic.
\end{description}
\end{prop}

The proof of Theorem \ref{Gen_Chasles} is based upon
the following property described in \cite{Fe}.

\begin{prop} \label{tangency} Let $G_I$, $I=\{i_1,\ldots i_k\}$ be the Pl\"ucker coordinates of a $k$-plane passing through
the origin in ${\mathbb R}^{m}$. 
 The set of all such $k$-planes that are tangent to a nondegenerate cone $\{\langle x,B x\rangle =0\}$,
$B={\rm diag}(b_1,\dots b_m)$
is the intersection of the (non-oriented) Grassmannian $G(m,k)\subset \wedge^k {\mathbb R}^{m}$ with the quadric
\begin{equation} \label{cone0}
 \left \{ \sum_I |B|_I^I \,G_I^2 =0\right \}\subset \wedge^k {\mathbb R}^{m}, \qquad |B|_I^I =b_{i_1}\cdots b_{i_k}.
\end{equation}
\end{prop}

\noindent{\it Proof of Theorem} \ref{Gen_Chasles}. Let us now set
$m+r$ and consider the family of cones (\ref{conf0}). Let, as
above, $\bar\Sigma$ be a $2r$-plane in ${\mathbb R}^{n+r}$ spanned
by the columns of $\mathcal V$ and associated to a point $X,P$ on
$T^*V_{n,r}$, and let
$$
B={\rm diag}\left(\frac{1}{\lambda- a_1},\dots, \frac{1}{\lambda- a_n}, 1, \dots, 1\right).
$$
Then, in view of Proposition \ref{tangency} and the structure of $\mathcal V$, for a fixed $\lambda=\lambda^*$
the set of the $2r$-planes that are tangent to $\bar Q(\lambda^*)$ is described by the following quadratic equation
in terms of the Pl\"ucker coordinates of $\bar\Sigma$
\begin{gather*}
\sum_{I_r} \frac{ (e_1\wedge \cdots \wedge e_r )_{I_r}^2 } {(\lambda^* -a_{i_1})\cdots (\lambda^* -a_{i_r}) } +
\sum_{I_{r+1} } \frac{1 }{(\lambda^* -a_{i_1}) \cdots (\lambda^* -a_{i_{r+1}} )}
\sum_{i=1}^n (e_1\wedge \cdots \wedge e_r\wedge p_i)_{I_{r+1}}^2 \\
 + \sum_{I_{r+2}} \frac{ 1 }{(\lambda^* -a_{i_1})
\cdots (\lambda^* -a_{i_{r+2}})}\sum_{1\le i<j\le n} (e_1\wedge \cdots \wedge e_r\wedge p_i\wedge p_j)_{I_{r+2}}^2 + \cdots
 \\
+\sum_{I_{2r}} \frac{ 1 }{ (\lambda^* -a_{i_1})\cdots (\lambda^* -a_{i_{2r}} ) } \,
(e_1\wedge \cdots \wedge e_r\wedge p_1\wedge \cdots \wedge p_r)_{I_{2r}}^2 =0 \,  .
\end{gather*}
Due to (\ref{2r-integrals}), this coincides with the equation ${\cal I}_{2r}(\lambda^*)=0$ up to multiplication
by $a(\lambda^*)$.

Since ${\cal I}_{2r}(\lambda)$ is also an invariant polynomial of degree $n-r$ in $\lambda$,
for a fixed plane $\bar\Sigma$ there exist precisely $n-r$ fixed cones of the family $\bar Q(\lambda)$ tangent to
$\bar\Sigma$. This establishes the theorem. \hfill$\Box$

\paragraph{Restriction to $\R^n$.} By analogy with the case $r=1$, one can
consider the restriction of family (\ref{conf0}) to the linear
subspace $\{y_1=\cdots=y_r=1\}$:
\begin{equation}
\label{conf} Q_r(\lambda) =\left\{
\frac{x_1^2}{a_1-\lambda}+\cdots+\frac{x_n^2}{a_n-\lambda} = r
\right \} =\mathfrak i^{-1}\left( \bar Q(\lambda) \cap
\{y_1=1,\dots,y_r=1\}\right),
\end{equation}
where $\mathfrak i: \R^n \to \R^{n+r}$
is the natural inclusion
$$
\mathfrak i(x_1,\dots,x_n)=(x_1,\dots,x_n,1,\dots,1).
$$
This gives a family of confocal quadrics in ${\mathbb R}^{n}$.

Further, the section of $\bar\Sigma$ by the subspace
$\{y_1=\cdots=y_r=1\}$ defines an affine $r$-plane
$\Sigma(t)=\mathfrak i^{-1}\left( \bar\Sigma \cap
\{y_1=\cdots=y_r=1\}\right)$ in ${\mathbb R}^{n}(x_1,\dots,x_n)$,
which is spanned by the orthogonal vectors $e_1, \cdots, e_r$ and
passes through the point $p_1+\cdots+p_r$. As a result, to a
generic solution $(X(t), P(t))$ of the Neumann system on $V_{n,r}$
one can uniquely associate the moving $r$-plane
$$
\Sigma(t)=p_1(t)+\cdots+p_r(t)+\Span\{e_1(t), \cdots, e_r(t)\}.
$$
In contrast to the case $r=1$, due to dimensional reasons, for $r>1$ the $r$-plane $\Sigma(t)$ is not necessarily
tangent to the quadrics $Q_r(c_1), \dots, Q_r(c_{n-r})$. More precisely, since
$$
d\,\mathfrak i\left(T_{(x_1,\dots,x_n)}
Q(\lambda)\right)=T_{\mathfrak i(x_1,\dots,x_n)}\bar
Q(\lambda)\cap \{y_1=1,\dots,y_r=1\},
$$
the tangency of $\bar\Sigma(t)$ and $\bar Q(c_i)$, for a fixed
$t$, either implies the tangency of the corresponding affine
$r$-plane $\Sigma(t)$ and the quadric $Q_r(c_i)$, or $\Sigma (t)$ does not intersect $Q_r(c_i)$.
As a result, one cannot formulate a natural generalization of the
Chasles theorem in ${\mathbb R}^{n}$ that involves this $r$-plane.

Another feature of the case $r>1$ is that,
although the first integrals given by the polynomial ${\cal I}_{2r}(\lambda)$ are invariant with respect to the right $SO(r)$-action on $(X,P)$, the $2r$-plane $\bar\Sigma$ and $r$-plane $\Sigma$ do not
have this property. Thus, a generic polynomial
${\cal I}_{2r}(\lambda)$ corresponds to a whole family of $2r$-planes ($r$-planes, respectively)
that are tangent to the same set of confocal cones and is obtained
as the orbit of $\bar\Sigma$ ($\Sigma$, respectively) under the right $SO(r)$-action.

Then, it natural to replace $\bar\Sigma$ by the moving cylinder $\bar\Delta(t)$,
the union of $2r$-planes $\bar\Sigma(X(t)B,P(t)B)$ spanned by the
columns of the $2r\times (n+r)$ matrices
$$
\mathcal V(X(t)B,P(t)B), \qquad B\in SO(r),
$$
where $\mathcal V(X,P)$ is given by (\ref{VV}). The cylinder
$\bar\Delta(t)$ is $SO(r)$-invariant and, due to the construction, is tangent simultaneously to $n-r$ fixed confocal
cones $\bar Q(c_1), \dots, \bar Q(c_{n-r})$.

Next, the section of $\bar\Delta(t)$ by the subspace $\{y_1=\cdots=y_r=1\}$ defines the moving
$(2r-1)$-dimensional cylinder
$$
\Delta (t)=\left\{\sum_{i,j} B_{i,j} p_i(t) \, \vert \, B\in SO(r) \right\}+\Span\{e_1(t), \cdots, e_r(t)\},
$$
which is now an appropriate object for the second generalization of the Chasles theorem:

\begin{thm} \label{Gen_Chasles2}
Let the $(2r-1)$-dimensional cylinder $\Delta(t)\subset {\mathbb
R}^n$ be associated to a generic solution $(X(t), P(t))$ of the
Neumann systems (\ref{N0_XP2}) or (\ref{NE_XP}) on $V_{n,r}$ as
described above. Then $\Delta(t)$ is tangent simultaneously to
$n-r$ fixed confocal quadrics $Q_r(c_1), \dots, Q_r(c_{n-r})$ of
the confocal family (\ref{conf0}), where $c_1,\dots, c_{n-r}$ are
the roots of the invariant polynomial ${\cal I}_{2r}(\lambda)$.
\end{thm}

\noindent{\it Proof.} First, note that the
plane $\bar\Sigma(X(t)B,P(t)B)$ can be obtained from
$\bar\Sigma(X(t),P(t))$ by rotating it in the coordinates
$y_1,\dots,y_r$ by the matrix $B^{-1}$.
That is, the cylinder $\bar\Delta(t)$ can be regarded as the orbit of
$\bar\Sigma(X(t),P(t))$ with respect to the $SO(r)$-action in the
coordinates $y_1,\dots,y_r$. This property is related to
the $SO(r)$-symmetry of the cones $\bar Q(\lambda)$ in
$y_1,\dots,y_r$.

Indeed, the $2r\times (n+r)$ matrices
\begin{equation*}
{\mathcal V}(XB,PB) \quad \mbox{and} \quad
 {\mathcal V}(XB,PB) \begin{pmatrix} B^{-1} & 0 \\ 0 & B^{-1} \end{pmatrix}=
\begin{pmatrix}  e_1 & \cdots & e_r & p_1 & \cdots & p_r \\
  0 & \cdots & 0 &   & & \\
\vdots &  & \vdots & & B^{-1} & \\
0   &  \cdots & 0 & &&
\end{pmatrix}
\label{VVB}
\end{equation*}
define the same $2r$-plane $\bar\Sigma(X(t)B,P(t)B)$, whereas the second matrix
is obtained from ${\mathcal V}(X,P)$ by left multiplication by
the block matrix $\diag(\mathbf{I}_{n-r}, B^{-1})$.

Now let $l(t)$ be the line along which the plane $\bar\Sigma(X(t),P(t))$ is tangent to the cone $\bar Q(c_i)$,
$$
l=l(t)=\Span\{v(t)=(v_1, v_2, \dots,v_{n+r} )^T\} \subset \bar\Sigma(X(t),P(t))\subset \R^{n+r}
$$
and $l_B(t)=\Span\{ v_B(t)\}$ be the tangency line of the rotated plane $\bar\Sigma(X(t)B,P(t)B)$ and $\bar Q(c_i)$.
Due to the above observation,
$v_B(t) =\diag(\mathbf{I}_{n}, B^{-1}) v(t)$.

One can always find $B^*\in SO(r)$ depending on $i$ and $t$,
such that the last $r$ coordinates of $v_{B^*}(t)$
are equal. Then the $r$-plane
$$
\Sigma_{B^*}(t)= \mathfrak i^{-1}\left(\bar\Sigma(X(t)B^*,P(t)B^*)\cap
\{y_1=\cdots=y_r=1\}\right)= \sum_{i,j} B_{i,j}^* p_i(t) +\Span\{e_1, \cdots, e_r\}
$$
is tangent to the quadric $Q_r(c_i)$ at the point
$\mathfrak i^{-1}\left(l_{B^*}(t) \cap \{y_1=\cdots=y_r=1\}\right)$. Since for any $i$ and $t$,
$\Sigma_{B^*}(t)$ is a subspace of the cylinder $\Delta(t)$, we arrive at the statement of the
theorem. \hfill $\Box$

\paragraph{Chasles Theorem for Manakov Flows.}
Similar statement holds for the geodesic flows of submersion
metrics defined by Manakov operators (see also Theorem 12 in \cite{Fe}).

Suppose $2r < n$. The dual Lax pair for the flow with Manakov operator (\ref{Manakov*}) given in Theorem
\ref{man-dual} gives the set of commuting integrals ${\mathcal
J}_{2l}(\lambda)$, the coefficients with term
$w^{2r-2l}$ in the expression $\vert a(\lambda) \mathcal
L^*_{man}(\lambda)-w{\bf I}_n\vert$. In particular,
$$
{\mathcal J}_{2r}(\lambda)= \sum_{I_{2r}} \frac{ a(\lambda) }{
(\lambda-a_{i_1})\cdots (\lambda-a_{i_{2r}} ) }
\,|\Phi|^{I_{2r}}_{I_{2r}}
$$
is a polynomial of degree $n-2r$ in $\lambda$.

Consider the following family of confocal cones in ${\mathbb
R}^{n}$:
$$
Q_0(\lambda) =\left\{ \frac{x_1^2}{a_1-\lambda}+\cdots+
\frac{x_n^2}{a_n-\lambda}=0 \right \}, \quad \lambda\in {\mathbb
R}.
$$

Repeating the arguments of Theorem \ref{Gen_Chasles}, and using
the fact that the Manakov geodesic flows with different choices of
the matrix $B$ in (\ref{Manakov}) are quasi-periodic motions over
the {\it same} isotropic toric foliation of $T^*V_{n,r}$, we get:

\begin{thm} \label{Chasles-Manakov}
Let the $2r$-plane
$$
\Sigma(t)=\Span\{e_1(t),\dots,e_r(t),p_1(t),\dots,p_r(t)\} \subset \R^n
$$
be associated to a generic solution $(X(t), P(t))$ of the geodesic flow (\ref{dot_XP2})
given by the Manakov operator (\ref{Manakov}).
Then $\Sigma(t)$ is tangent
simultaneously to $n-2r$ fixed confocal cones $Q_0(c_1), \dots,
Q_0(c_{n-2r})$, where $c_1,\dots, c_{n-2r}$ are the roots of the
invariant polynomial ${\mathcal J}_{2r}(\lambda)$.
\end{thm}

\begin{remark}{\rm
Since all objects in theorems \ref{Gen_Chasles2} and
\ref{Chasles-Manakov} are right $SO(r)$-invariant they are also
valid for the Neumann system on the oriented Grassmannian variety
$G_{n,r}$ as well as for the geodesic flows on  $G_{n,r}$ obtained
by submersion from the Manakov flows. }\end{remark}

\paragraph{The case of the Poisson sphere.}
As an illustrative example, consider the case $r=1$ and Manakov operator (\ref{Manakov}) with $B=-A^{-1}$.
Then the submersion metric on $S^{n-1}$ takes the following form (see \cite{Br})
\begin{equation}
ds^2=\frac{1}{\langle A^{-1}e,e\rangle} \sum_{i=1}^n a_i d e_i^2. \label{sp1}
\end{equation}
In the elliptic coordinates the metric (\ref{sp1}) is of the St\"ackel type, and its geodesic flow
\begin{eqnarray*}
&& \dot e=(A^{-1}e,e)A^{-1}p-(A^{-1}e,p)A^{-1}e \label{P_q},\\
&& \dot p=(A^{-1}e,p)A^{-1}p-(A^{-1}p,p)A^{-1}e \label{P_p}
\end{eqnarray*}
is completely integrable \cite{Br}. For $n=3$,  the metric (\ref{sp1})
is proportional to the metric on the Poisson sphere $S^{2}$, i.e.,
to the metric obtained after $SO(2)$ reduction of the free rigid
body motion around a fixed point with the inertia tensor $I=A^{-1}$.

From Theorem \ref{Chasles-Manakov}, we get

\begin{cor}
Let $e(t)$ be a geodesic line of the Poisson sphere metric (\ref{sp1}). Then the  moving $2$-plane
$$
\Sigma(t)=\Span\{ e(t), \, p(t)\}=\Span\{ e(t), \,A\dot e(t)\}
$$
is tangent to $n-2$ fixed confocal
quadrics of the family $Q_0(\lambda)$.
\end{cor}

\section{Appendix 1. Magnetic Neumann Systems}

\paragraph{The Neumann system on $S^{n-1}$ in presence of the Yang--Mills fields.}
We now go back to the reduction of the Neumann system to
Grassmannians described in Section \ref{grassmannian} and consider in detail the case $r-1$.

Apart from the complete reduction of the Neumann system onto
$\Psi^{-1}(0)/SO(n-1)\cong T^*S^{n-1}$, it is also convenient to
describe partially reduced flows for non-zero values of the
momentum $\Psi$, that is, the flows on
\begin{equation}
(T^*V_{n,n-1})/SO(n-1) \cong so(n)\times S^{n-1}, \label{mag0}
\end{equation}
as well as the reduced flows on the symplectic leaves
$$
\Psi^{-1}(\eta)/SO(n-1)_\eta \cong \Psi^{-1}(\mathcal
O_\eta)/SO(n-1)\subset (T^*V_{n,n-1})/SO(n-1),
$$
where $\mathcal O_\eta$ is the adjoint orbit of $\eta$. It is
known that for $\eta\ne 0$, the quotients $\Psi^{-1}(\mathcal
O_\eta)/SO(n-1)$ are diffeomorphic to $\mathcal O_\eta$-bundles
over (co)tangent bundle of the sphere $S^{n-1}$ and that the
reduced systems are natural mechanical systems with the influence
of the Yang--Mills fields (see, e.g., \cite{GS}, Chapter III, which
provides a detailed geometrical analysis of such systems).

We shall derive the reduced equations and Yang--Mills fields directly
from the equations of motion. Namely, for a matrix $\xi\in so(n)$
and an unit vector $e_n\in S^{n-1}$, define projections
$$
\pr_{e_n}(\xi)=\xi e_n\otimes e_n+ e_n\otimes e_n \xi, \quad
\pr_{{e_n}^\perp}(\xi)=\xi-\pr_{e_n}(\xi)
$$
with respect to the orthogonal decomposition
\begin{equation}
so(n)=\{e_n \wedge \R^n\}\oplus \{e_n \wedge \R^n\}^{\perp}.
\label{OD}
\end{equation}

Note that $\{e_n \wedge \R^n\}^{\perp}\cong so(n-1)$ and $\{e_n
\wedge \R^n\}$ can be naturally identified with the tangent space
$T_{e_n} S^{n-1}$. Therefore the reduced phase space
$T^*V_{n,n-1}/SO(n-1)$ is also represented as a $so(n-1)$-bundle
over $T S^{n-1}$ that we shall denote by $so(n-1)\times_s T
S^{n-1}$. There is the natural inclusion
$$
so(n-1)\times_s T S^{n-1} \subset so(n)\times TS^{n-1}: \quad
(\xi,e_n,\dot e_n) \in so(n-1)\times_s TS^{n-1} \, \Leftrightarrow
\, \pr_{e_n}(\xi)=0.
$$

\begin{prop} \label{YM}  \begin{description}
\item{(i)} The reduced Neumann system on the quotient variety
(\ref{mag0}) has the form
\begin{equation}
\dot \Omega=[A,e_n\otimes e_n], \quad \dot e_n=\Omega e_n , \qquad
\Omega\in so(n). \label{Euler**}
\end{equation}
\item{(ii)} The second derivative of the vector $e_n$ is
\begin{equation}
\ddot e_n=A e_n-\left( ( e_n,A e_n )+( \dot e_n,\dot e_n )
\right)e_n+F_\Omega\, \dot e_n, \label{mag6}
\end{equation}
where $F_\Omega=\pr_{{e_n}^\perp}(\Omega)=\Omega-\Phi(e_n,\dot
e_n)$ and $\Phi(e_n,\dot e_n)=\dot e_n \wedge e_n$ is the standard
$SO(n)$-momentum mapping on $TS^{n-1}$.
\end{description}
\end{prop}

If we restrict the flow to the invariant submanifold
$\Psi^{-1}(\mathcal O_\eta)/SO(r)$, then $(F_\Omega,e_n,\dot e_n)$
ranges over the subbundle $\mathcal O_\eta \times_s T S^{n-1}\subset
so(n-1)\times_s T S^{n-1}$ obtained by replacing fibers $so(n-1)$
by the adjoint orbits $\mathcal O_\eta \subset so(n-1)$. In particular, for
$\eta=0$ we recover (\ref{neum-gras**}). For $\eta\ne 0$, the
additional term $F_\Omega\,\dot e_n$ can be interpreted as the
influence of the Yang--Mills field with the internal symmetry group
$SO(n-1)$ and charge type $\mathcal O_\eta$ (\cite{GS}). In the
special case $n=3$, $r=2$, after identification of the Lie
algebras $(so(3), [\cdot,\cdot])$ and $(\R^3, \times)$,  equation
(\ref{mag6}) takes the form
\begin{equation*}
\ddot e_3=A e_3- \left( ( e_3,A e_3 )+ (\dot e_3,\dot e_3 )
\right) e_3+\epsilon \, e_3\times\dot e_3,
\end{equation*}
which describes the motion of the particle with the charge
$\epsilon$ on $S^2$ in the magnetic monopole field. Here
$\eta=\epsilon E_1\wedge E_2$ and $F_\Omega\,\dot e_3=\epsilon \,
e_3\times\dot e_3$ represents the Lorentz force of the magnetic
monopole.

Since we already proved the completeness of the commuting
integrals (\ref{perelomov}) on a generic symplectic leaf within
(\ref{mag0}), we arrive at the following statement.

\begin{thm}
The Neumann system perturbed by the Yang--Mills
field, i.e, the restriction of (\ref{Euler**}) to $\mathcal O_\eta
\times_s T S^{n-1}$, is completely integrable for a generic value
$\eta\in so(n-1)$.
\end{thm}

\noindent{\it Proof of Proposition \ref{YM}.} Let us identify
$V_{n,n-1}$ and $SO(n)$ via (\ref{iden1}) and consider $SO(n)$ as
the configuration space of the rigid body moving around a fixed
point. Then the vectors $e_1,\dots,e_n$ are fixed in the body, the
matrix $\mathcal X=(e_1,\dots,e_n)$ maps the fixed frame to the
frame attached to the body and
$$
M=\mathcal P\mathcal X^T-\mathcal X\mathcal P^T
$$
plays the role of the angular momentum of the body {\it in the
space frame}. Here we denoted the $n\times n$ momentum matrix by
$\mathcal P$.

From the identity $ e_n\otimes e_n=\mathbf{I}_n-\sum_{i=1}^{n-1}
e_i\otimes e_i $ we see that, after identification (\ref{iden1}),
the Neumann system with the normal metric on $V_{n,n-1}$
corresponds to the motion of the rigid body with the Hamiltonian
\begin{equation}
H_{CP}=\frac12\langle M,M\rangle+\frac12\left(\tr A- ( e_n, A e_n
)\right). \label{mag*}
\end{equation}

This is a special, symmetric case of the Clebsch--Perelomov rigid
body problem: the inertia operator of the body is the
identity on the Lie algebra $so(n)$ (see \cite{Pe}). Therefore the
 angular velocity in {\it the space frame} $ \Omega=\dot {\mathcal
X}\cdot {\mathcal X}^{-1} $ and the angular velocity in {\it the body
frame} $\omega={\mathcal X}^{-1}\dot{\mathcal X}$ are equal to the
angular momentum in the space frame $M$ and to the angular
momentum in the body frame $m=\mathcal X^{-1}M\mathcal X$ respectively.

The Hamiltonian (\ref{mag*}) is invariant with respect to rotations in
${\mathbb R}^{n-1}=\text{span} (e_1,\dots,e_{n-1})$, i.e., it is right $SO(n-1)$-invariant.
In the right trivialization, the motion of the body is described by the Euler--Poincar\'e equations
\begin{equation}
\dot \Omega=[A,e_n\otimes e_n] \label{Euler*}
\end{equation}
together with the Poisson equations
\begin{equation}
\dot e_i=\Omega e_i, \qquad i=1,\dots,n. \label{Poisson}
\end{equation}
Whence it is clear that the reduced flow is given by
(\ref{Euler*}) and the last Poisson equation
\begin{equation}
\dot e_n=\Omega e_n.\label{mag2}
\end{equation}

To prove the second assertion of Proposition \ref{YM}, we rewrite
equations (\ref{Euler**}) with respect to the orthogonal
decomposition (\ref{OD}). Then, from (\ref{mag2}), we have $\dot
e_n=\pr_{e_n}(\Omega)e_n$ and
\begin{equation}
\pr_{e_n}(\Omega)=\dot e_n \wedge e_n =\Phi(e_n,\dot e_n).
\label{mag3}
\end{equation}

In view of (\ref{Euler*}) and (\ref{mag3}), the time derivation of (\ref{mag2}) reads
\begin{eqnarray*}
\ddot e_n &=& \dot\Omega e_n+\Omega\dot e_n =\dot \Omega e_n+\pr_{{e_n}}(\Omega)\dot e_n+\pr_{{e_n}^\perp}(\Omega)\dot e_n \\
   &=& (Ae_n\otimes e_n-e_n\otimes e_n A)e_n+(\dot e_n \wedge e_n)\dot e_n + \pr_{{e_n}^\perp}(\Omega)\dot e_n \\
&=& A e_n-\left( ( e_n,A e_n )+( \dot e_n,\dot e_n )
\right)e_n+F_\Omega\, \dot e_n,
\end{eqnarray*}
which concludes the proof. \hfill $\Box$

\paragraph{Symmetric Clebsch--Perelomov--Bogoyavlenski rigid body systems.}
A similar reduction can be made for the Neumann systems on
$V_{n,r}$ and $G_{n,r}$ by considering the motion of a symmetric rigid body
with the Hamiltonian
$$
H_{CPB}=\frac12\langle M,M\rangle+\frac12\sum_{i=1}^r\ ( e_i, A
e_i ) \, ,
$$
which corresponds to a special (symmetric) case of the Bogoyavlenski generalization of the
Clebsch--Perelomov system (see \cite{Bog}). Namely, in the space frame the motion is described by the
Euler--Poincar\'e equations
\begin{equation}
\dot \Omega=[e_1\otimes e_1 + \dots + e_r \otimes e_r, A]
\label{Euler}
\end{equation}
together with the Poisson equations (\ref{Poisson}). After
substitutions
$$
h=\Omega \quad  {\rm and} \quad  v=e_1\otimes e_1 + \dots + e_r
\otimes e_r
$$
they take the closed form (\ref{CPB}).

The system is right $SO(r)\times SO(n-r)$-invariant with the
momentum mapping
$$
\Psi=\Psi_{so(r)} + \Psi_{so(n-r)}, \quad
\Psi_{so(r)}=\pr_{so(r)}({\mathcal X}^{-1}\Omega{\mathcal X}),
\quad \Psi_{so(n-r)}=\pr_{so(n-r)}({\mathcal
X}^{-1}\Omega{\mathcal X}),
$$
where
$$
so(r)=\Span\{E_i\wedge E_j, \, 1 \le i < j \le r\},\quad
so(n-r)=\Span\{E_i\wedge E_j, \, r+1 \le i < j \le n\}.
$$

The reductions of (\ref{Euler}), (\ref{Poisson}) to
$\Psi_{so(n-r)}^{-1}(\mathcal O_{\eta_{so(n-r)}})/SO(n-r)$ and to
$$
\Psi^{-1}(\mathcal O_{\eta_{so(r)}}\times\mathcal O_{\eta_{so(n-r)}})/SO(r)\times SO(n-r)
$$
lead to the Neumann systems on the Stiefel variety $V_{n,r}$ and,
respectively, to the oriented Grassmannian variety $G_{n,r}$ under
the influence of the Yang--Mills fields. In particular, if
$\eta_{so(r)}=0$ ($\eta_{so(r)}=\eta_{so(n-r)}=0$), we get the
Neumann system with the {\it normal metric} on $V_{n,r}$ (respectively,
the Neumann system on $G_{n,r}$).

\paragraph{Magnetic Neumann flows on $V_{n,2}$ and $G_{n,2}$.} The adjoint orbits in $so(2)$ are points, so
a symplectic reduced space
$\Psi^{-1}((\eta_{so(2)},0))/SO(2)\times SO(n-2)$ is diffeomorphic
to $T^*G_{n,2}$.
For $ \eta_{so(2)}=\epsilon\, E_1 \wedge E_2\ne
0,$ it represents the magnetic cotangent bundle $T^*G_{n,2}$:
the canonical symplectic structure of $T^*G_{n,2}$ is ''twisted'' by
adding the magnetic form, which is exactly Kirillov--Konstant
symplectic form on $G_{n,2}$ multiplied by $\epsilon$ (for more
details see, e.g.,  \cite{Ku, BJ4}).

As above, for us it is convenient to write the equations in the
Euler--Lagrange form. Due to the presence of the magnetic form,
the tangent bundle momentum mapping of the left $SO(n)$-action is
modified by adding the term $\epsilon\, e_1 \wedge e_2$ (see
\cite{Ef, BJ4})
\begin{equation}
\Phi_\epsilon =\dot e_1 \wedge e_1+ \dot e_2 \wedge e_2+\epsilon\,
e_1 \wedge e_2= [ e_1 \wedge e_2, \dot e_1 \wedge e_2+ e_1 \wedge
\dot e_2]+\epsilon\, e_1 \wedge e_2 \, , \label{mag-mom-grass}
\end{equation}
whereas the right-hand side of the Neumann system
(\ref{neum-gras}) is modified by adding the term
$\epsilon\,\Phi_0$ (see \cite{BJ4})
\begin{eqnarray}
\frac{d^2}{dt^2}\left(e_1\wedge e_2\right) &=&
-Ae_1 \wedge e_2-e_1\wedge A e_2 \nonumber\\
&&+\left((e_1,Ae_1 )- (\dot e_1,\dot e_1) + ( e_2,Ae_2 )- (\dot e_2,\dot e_2)\right) e_1 \wedge e_2 \label{neum-gras-mag}\\
&&+2 \dot e_1\wedge \dot e_2+\epsilon\, [ e_1 \wedge e_2, \dot e_1
\wedge e_2+ e_1 \wedge \dot e_2].\nonumber
\end{eqnarray}
Here $e_1,e_2,\dot e_1,\dot e_2$ satisfy the conditions
(\ref{gras-velocities}). (Clearly, a similar symplectic reduction with
a magnetic term can also be applied on the Stiefel variety $V_{n,n-2}$ (see \cite{RS}).)

To construct the magnetic Neumann flows on $T^*V_{n,2}$, consider closed 2-form
$$
\omega_{mag}=de_1 \wedge de_2=\sum_{i=1}^n de_1^i \wedge de_2^j
$$
restricted to $V_{n,2}$. Let $\pi: T^*V_{n,2}\to V_{n,2}$ be the
canonical projection and define the symplectic form
\begin{equation}
\omega_\epsilon=\omega+\epsilon\,\pi^*\omega_{mag},
\label{twisted}
\end{equation}
where $\omega$ is the canonical form on $T^*V_{n,2}$ (see Section
\ref{first}).

The following two propositions can be verified by straightforward
computations.

\begin{prop}
The left $SO(n)$-action on $(T^*V_{n,2}, \omega_\epsilon)$ is
Hamiltonian with the momentum mapping given by
\begin{equation}
\Phi_{\epsilon}=\Phi+\epsilon \, e_1 \wedge e_2=p_1 \wedge e_1+
p_2 \wedge e_2 + \epsilon\,  e_1 \wedge e_2. \label{mag-mom-stief}
\end{equation}
\end{prop}

\begin{prop}
The Hamiltonian equations defined by the Hamiltonians of the
Neumann systems with the Euclidean metric and the Normal metric
with respect to the symplectic structure (\ref{twisted}) read
\begin{eqnarray} && \dot e_1=p_1, \nonumber\\ && \dot e_2=p_2, \label{mag-euk} \\
&& \dot p_1=-Ae_1+\left((e_1,A e_1)-(p_1,p_1)-\epsilon (e_1,p_2)\right) e_1 + \left((e_1,A e_2)-(p_1,p_2)\right) e_2 + \epsilon\, p_2, \nonumber\\
&& \dot p_2=-Ae_2+\left((e_1,A e_2)-(p_1,p_2)\right) e_1 +
\left((e_2,A e_2)-(p_2,p_2)+\epsilon(e_2,p_1)\right) e_2 -
\epsilon \,p_1 \nonumber
\end{eqnarray}
and, respectively,
\begin{eqnarray}
&& \dot e_1=\Phi_0 e_1=p_1-(e_1,p_2) e_2, \nonumber\\
&& \dot e_2=\Phi_0 e_2=p_2-(e_2,p_1) e_1, \label{mag-nor}\\
&& \dot p_1=\Phi_0 p_1- Ae_1 + \left( (e_1,Ae_1)-\epsilon(e_1,\Phi_0 e_2)\right) e_1+(e_1,Ae_2)e_2+ \epsilon\, \Phi_0 e_2, \nonumber\\
&& \dot p_2=\Phi_0 p_2- Ae_2 + (e_2, Ae_1) e_1+ \left(
(e_2,Ae_2)+\epsilon(e_2,\Phi_0 e_1)\right) e_2- \epsilon\, \Phi_0
e_1 \, . \nonumber
\end{eqnarray}
\end{prop}

Equations (\ref{mag-euk}), (\ref{mag-nor}) are right
$SO(2)$-invariant and have integral
$\Psi_{12}=(e_1,p_2)-(e_2,p_1)$. The magnetic Neumann system
(\ref{neum-gras-mag}) can be also seen as a reduction of the system
(\ref{mag-euk}), or (\ref{mag-nor}) with respect to the right
$SO(2)$-action.

Furthermore, for the systems (\ref{neum-gras-mag}), (\ref{mag-euk}),
(\ref{mag-nor}) the relation (\ref{prep}) still holds in the form
$$
\frac{d}{dt} \Phi_\epsilon=[e_1 \otimes e_1+e_2 \otimes e_2, A],
\quad \frac{d}{dt}(e_1 \otimes e_1+e_2 \otimes
e_2)=[\Phi_\epsilon, e_1 \otimes e_1+e_2 \otimes e_2], $$ which
implies the Lax representation (\ref{LA1}), where instead of the
momentum mapping $\Phi$ one should use $\Phi_\epsilon$ in
\eqref{mag-mom-stief}.

\begin{thm}
The magnetic Neumann systems  (\ref{neum-gras-mag}),
(\ref{mag-euk}) and  (\ref{mag-nor}) are completely integrable in
the commutative sense with respect to the twisted symplectic
structures described above.
\end{thm}

The proof is a simple modification of those of Theorems \ref{NEUMANN} and \ref{NEUMANN2}.

\section{Appendix 2. Rank $r$ Double, Coupled and Neumann Systems on Complex Stiefel Manifolds}

In this section we briefly consider several natural
generalizations of the Neumann flows on Stiefel varieties. We present their equations of motion and Lax
representations, however a complete verification of the
integrability is out of the scope of this paper.

\paragraph{Rank $r$ double Neumann system.}
In \cite{Su} Suris introduced the double Neumann system describing the
motion of 2 points $x,y\in {\mathbb R}^n$ which interact via the
bilinear potential $(x,Ay)/2$ under the constraint $(x,y)=1$.

We consider {\it rank $r$ double Neumann system} defined by the
Lagrangian function
\begin{equation}
L(X,Y,\dot X,\dot{Y})=\tr(\dot X^T \dot{Y}) -\tr(X^T A Y),
\label{D-Lag}
\end{equation}
where the $n \times r$ matrices $X,Y\in M_{n,r}(\mathbb R)$ are
subject to the constraints
\begin{equation}
{X}^T {Y}={\bf I}_r. \label{D0}
\end{equation}

The corresponding Euler--Lagrange equations with $r\times r$ matrix multipliers
read:\footnote{For simplicity here we only give the Lagrangian description.}
\begin{equation}
 \ddot{X}=-AX+X\Lambda^T, \quad \ddot{Y}=-A
Y+Y\Lambda, \label{D-eq}\end{equation} where
\begin{equation}
\Lambda=X^T A Y-\dot{X}^T \dot Y. \label{D-L}
\end{equation}

The rank $r$ double Neumann system is an extension of the Neumann
system on $V_{n,r}$ with the Euclidean metric: if $(X(t),Y(t))$ is a
solution of the system (\ref{D-eq}) with the initial conditions
$X=Y$, $\dot X=\dot{Y}$, then $(X(t),P(t))=(X(t),\dot X(t))$ is a
solution of (\ref{NE_XP}). The Lax representation \eqref{LA3}
extends as follows.

\begin{thm} \label{dual-D}
The equations (\ref{D-eq}) imply the following $2r\times 2r$
matrix Lax pair with the parameter $\lambda$
\begin{gather} \label{LA3*D}
\frac{d}{dt} \mathcal L(\lambda)=[\mathcal L(\lambda),\mathcal
A(\lambda)] \, , \\
\mathcal {L}(\lambda)=\begin{pmatrix}
-X^T(\lambda \mathbf{I}_n-A)^{-1} \dot{Y} & -X^T(\mathbf{I}_n-\lambda A)^{-1} Y \\
\mathbf{I}_r+\dot X^T(\mathbf{I}_n-\lambda A)^{-1} \dot{Y} & \dot
X^T(\lambda \mathbf{I}_n-A)^{-1} Y \end{pmatrix}, \quad
\mathcal A(\lambda)=\begin{pmatrix} 0 & \mathbf{I}_r \\
\Lambda-\lambda \mathbf{I}_r & 0 \end{pmatrix}\, ,  \notag 
\end{gather}
$\Lambda$ being given by (\ref{D-L}). \end{thm}

Apart from the integrals provided by the Lax matrix $L(\lambda)$,
the equations also possess the matrix integral $X^T \dot Y - Y^T
\dot X$ associated to the $GL(n,\mathbb R)$-symmetry
$$
(X,Y,\dot X,\dot Y) \longmapsto (X R^T,Y R^{-1},\dot X R^T, \dot Y
R^{-1}), \qquad R\in GL(n,\R).
$$

\paragraph{Coupled Neumann system on $V_{n,r}$.}
This systems generalizes the motion of 2 points $x,y$ on the unit
sphere $S^{n-1}\subset {\mathbb R}^n$ that interact via the bilinear potential
$(x,Ay)/2$. The latter system was introduced in \cite{RS} (see also \cite{Su}).

Namely, let matrices $X,Y\in M_{n,r}(\mathbb R)$ define two
points on $V_{n,r}$ and let $P,Q\in M_{n,r}(\mathbb R)$ represent
their momenta such that
$$
 X^T P+ P^T X= 0, \quad  Y^T Q+ Q^T Y=0 .
$$
Assume that the evolution of $X, Y$ is described by the Hamiltonian
\begin{gather} \label{C-neumann}
H= T_\kappa + \tr(X^TA Y), \\
T_\kappa = \frac12\tr(P^TP)-\left(\frac12+\kappa\right)\tr((X^T
P)^2)+\frac12\tr(Q^TQ)-\left(\frac12+\kappa\right)\tr((Y^T Q)^2),
\notag
\end{gather}
where $T_\kappa$ is the kinetic energy of the points defined by an
$SO(n)$-invariant metric on $V_{n,r}$, which depends on the
parameter $\kappa$. As above (see Section 3), for $\kappa= -1/2$ we have the
Euclidean  metric and for $\kappa =0$ the normal one.
(Note that in the case $r=1$ we have $X^T P=Y^T Q=0$, and all the above metrics coincide.)

The Hamilton equations with multipliers have the form
\begin{equation}
\begin{aligned}
& \dot X=P-(1+2\kappa) X P^T X, \\
& \dot P=-AY+(1+2\kappa) PX^T P+X\Lambda,\\
& \dot Y=Q-(1+2\kappa) Y Q^T Y, \\
& \dot Q=-AX+(1+2\kappa)QY^T Q+Q\Pi,
\end{aligned}\label{C-eq}
\end{equation}
with
\begin{equation}
\Lambda=\frac12(X^T A Y+Y^TAX)-P^T P, \quad \Pi=\frac12(X^T A
Y+Y^TAX)-Q^T Q. \label{C-L}
\end{equation}

Borrowing the terminology of \cite{Su}, we call \eqref{C-eq} the
{\it $r$-coupled Neumann systems} on $V_{n,r}$. They are invariant
with respect to the right diagonal $SO(r)$-action on the product
$T^*V_{n,r}\times T^*V_{n,r}$, and the corresponding matrix
momentum $X^TP-P^TX+ Y^TQ-Q^TY$ is preserved along their flows.

The book \cite{RS} presented a ''big'' Lax pair of the
$r$-coupled systems with the $so(n,n)$-matrices depending on
parameter $\nu$
\begin{gather}\label{so(n,n)}
\dot L(\nu) =[L(\nu), A(\nu)], \notag \\
 L(\nu) = \begin{pmatrix} {\bf 0} & A \\
                         A & {\bf 0} \end{pmatrix}\nu +
\begin{pmatrix} P X^T-X P^T & {\bf 0} \\
             {\bf 0} & Q Y^T-Y Q^T \end{pmatrix} +
\begin{pmatrix} {\bf 0} & X Y^T \\
                         Y X^T & {\bf 0} \end{pmatrix}\nu^{-1} , \\
    A(\nu) = L_+(\lambda)= \begin{pmatrix} {\bf 0} & A \\
                         A & {\bf 0} \end{pmatrix}\nu +
\begin{pmatrix} P X^T-X P^T & {\bf 0} \\
             {\bf 0} & Q Y^T-Y Q^T \end{pmatrix} , \notag
\end{gather}
${\bf 0}$ being zero $n\times n$ block. Curiously, this Lax pair
holds only for the case of the normal metric ($\kappa=0$), and not
for the Euclidean one, as one might expect and as happens in the case
of the Neumann flows on $T^*V_{n,r}$.

Below we also present the dual ''small'' Lax representation. Introduce
$r\times r$ matrix
$$
F_\lambda(X,Y)=X^T(A^2-\lambda^2)^{-1}Y.
$$

\begin{thm} \label{dual-DC}
The coupled Neumann system (\ref{C-eq}) with $\kappa=0$ admits the
following Lax pair with the spectral parameter $\lambda$
\begin{equation}
\frac{d}{dt} \mathcal L(\lambda)=[\mathcal L(\lambda),\mathcal
A(\lambda)]\label{LA3*C},
\end{equation}
$\mathcal L(\lambda), \mathcal A(\lambda)$ being $4r\times 4r$
matrices
\begin{eqnarray*}
&\mathcal {L}(\lambda)=\begin{pmatrix} \lambda F_\lambda(P,X) &
\lambda F_\lambda(P,P) & F_\lambda(AP,Y) & -{\bf
I}_r +F_\lambda(AP,Q) \\
-\lambda F_\lambda(X,X) & -\lambda F_\lambda(X,P) &
-F_\lambda(AX,Y) & -F_\lambda(AX,Q) \\
F_\lambda(AQ,X) & -{\bf I}_r+F_\lambda(A Q,P) &
\lambda F_\lambda(Q,Y) & \lambda F_\lambda(Q,Q) \\
-F_\lambda(AY,X) & -F_\lambda(AY,P) & -\lambda F_\lambda(Y,Y) &
-\lambda F_\lambda(Y,Q)
 \end{pmatrix},\\
& \mathcal A(\lambda)=\begin{pmatrix} X^T P & \Lambda & 0 & -\lambda \mathbf{I}_r \\
\mathbf{I}_r & X^T P & 0 & 0 \\
0 & -\lambda \mathbf{I}_r & Y^T Q & \Pi \\
0 & 0 & \mathbf{I}_r & Y^T Q
\end{pmatrix}\, , \label{LA4*C}
\end{eqnarray*}
and $\Lambda$, $\Pi$ are given by (\ref{C-L}). \end{thm}

Again, the proof is straightforward. For $r=1$ the Lax pair
\eqref{LA3*C} was given by Suris in \cite{Su}.

It is still not clear whether the $r$-coupled Neumann system with
the Euclidean metric admits a Lax representation. However, we can
consider the following perturbation of the Hamiltonian \eqref{C-neumann}:
$$
H_{\kappa}= T_\kappa + \tr(X^TA Y)-2\kappa\tr(X^TP\, Y^TQ).
$$
Then the corresponding flows imply the Lax
representation with matrices \eqref{so(n,n)} for any $\kappa$,
while the Lax representation \eqref{LA4*C} holds with $X^TP$ and
$Y^T Q$ on the diagonal of $\mathcal A(\lambda)$ replaced by
$(1+2\kappa)X^TP+2\kappa Y^TQ$ and  $2\kappa X^TP+(1+2\kappa)
Y^TQ$, respectively. In particular, by taking $\kappa=-1/2$, we
get the Lax representation of an $r$-coupled Neumann system with
the Euclidean metric and with an additional interacting term
$\tr(X^TP \, Y^TQ)$.

\paragraph{Complex Stiefel manifolds.}
The Neumann systems and geodesic flows on $V_{n,r}$ can be
extended to the complex Stiefel varieties $W_{n,r}$ as well.
Recall that $W_{n,r}$ is the space of $r$ ordered orthogonal
vectors $(z_1,\dots,z_n)$ in $\mathbb{C}^n$ endowed with the
standard Hermitian metric, or equivalently, the set of $n\times r$
matrices $ Z\in M_{n,r}(\mathbb C)$ satisfying
\begin{equation}
{\bar Z}^T {Z}={\bf I}_r  \label{W0}
\end{equation}
(see, e.g., \cite{KN}). The variety $W_{n,r}$ can also be identified with the
homogeneous space of the unitary group: $W_{n,r} \cong
U(n)/U(n-r)$. The real Stiefel variety $V_{n,r}$ is thus a submanifold
of $W_{n,r}$ given by the condition $Z=\bar Z$.

While the idea of integrable geodesic flows on $W_{n,r}$
follows from the general construction given for compact homogeneous
spaces \cite{BJ1, BJ2, BJ3}, to our knowledge, potential
systems on $W_{n,r}$ for $r>1$ were not studied yet.

We shall consider the Neumann system with the metric induced by
the {\it Hermitian metric} and defined by the Lagrangian function
\begin{equation}
L(Z,\bar Z,\dot Z,\dot{\bar Z})=\frac12\tr(\dot Z^T \dot{\bar Z})
-\frac12\tr(Z^T A \bar Z). \label{W-Lag}
\end{equation}
As above, the matrix $A$ is a real diagonal $n\times n$ matrix.

The Euler--Lagrange equations with multipliers read:
\begin{equation}
 \ddot{Z}=-AZ+Z\Lambda, \quad \ddot{\bar Z}=-A\bar
Z+\bar Z\bar\Lambda, \label{W-eq}
\end{equation}
where
\begin{equation} \Lambda=\bar Z^T A Z-\dot{\bar Z}^T \dot Z=\bar\Lambda^T. \label{W-L} \end{equation}

Then the Neumann system on $V_{n,r}$ with the Euclidean metric
(\ref{NE_XP}) can be regarded as a subsystem of (\ref{W-eq}): if
$Z(t)$ is its solution on the complex Stiefel variety $W_{n,r}$
with the initial conditions satisfying $Z=\bar Z$, $\dot
Z=\dot{\bar Z}$, then $(X(t),P(t))=(Z(t),\dot Z(t))$ is a solution
of the Neumann system (\ref{NE_XP}) on $V_{n,r}$, and vice versa.

We also have

\begin{thm} \label{dual-complex}
The Neumann system on the complex Stiefel manifold (\ref{W-eq}) imply
the $2r\times 2r$ matrix representation with the spectral parameter $\lambda$
\begin{gather}
\frac{d}{dt} \mathcal L(\lambda)=[\mathcal L(\lambda),\mathcal
A(\lambda)]\, , \label{LA3*} \\
\mathcal {L}(\lambda)=\begin{pmatrix}
-Z^T(\lambda \mathbf{I}_n-A)^{-1} \dot{\bar Z} & -Z^T(\mathbf{I}_n-\lambda A)^{-1} \bar Z \\
\mathbf{I}_r+\dot Z^T(\mathbf{I}_n-\lambda A)^{-1} \dot{\bar Z} &
\dot Z^T(\lambda \mathbf{I}_n-A)^{-1} \bar Z \end{pmatrix}, \quad
\mathcal A(\lambda)=\begin{pmatrix} 0 & \mathbf{I}_r \\
\bar \Lambda-\lambda \mathbf{I}_r & 0 \end{pmatrix}\, ,  \notag 
\end{gather}
with $\Lambda$ given by (\ref{W-L}). \end{thm}

The system is invariant with respect to a right $U(r)$-action and
the symmetry with respect to the left action of $U(1)^n$ defined
by
\begin{equation}
Z \longmapsto \diag(\rho_1,\dots,\rho_n)\cdot Z, \qquad \rho_i\in
U(1), \quad i=1,\dots,n. \label{u(1)}
\end{equation}
The symmetries imply the conservation of $u(n)$ and $u(1)^n$
momentum maps
$$
\psi_0=Z^T\dot{\bar Z}-\dot Z^T\bar Z \quad \text{and} \quad
\psi_j=\sum_{i=1}^r \dot z_i^j\bar z_i^j - z_i^j\dot{\bar z}_i^j,
\quad j=1,\dots,n,
$$
where $Z=(z_1,\dots,z_r)$.

Under the condition $\rho_1=\dots=\rho_n$, (\ref{u(1)}) defines
the left $U(1)$-action, which is free. Then we have a well defined
reduced Neumann flow on the quotient space ({\it complex
projective Stiefel variety}) $PW_{n,r}=W_{n,r}/U(1)$.

On the other
hand, the right $U(r)$-symmetry enables one to reduce the Neumann
system to the complex Grassmann variety $ G_\mathbb C(n,r) \cong
U(n)/U(r)\times U(n-r) $ of $r$-dimensional complex planes in
$\mathbb{C}^n$, or, in general, to $U(n)$-adjoint orbits $
\mathcal O=U(n)/U(k_1)\times \dots \times U(k_l)\times U(n-r)$,
where $k_1+k_2+\dots+k_l=r. $

\subsection*{Acknowledgments}
We are grateful to Yuri G. Nikonorov on useful discussions. The
research of B. J. was supported by the Serbian Ministry of
Science, Project 174020, Geometry and Topology of Manifolds,
Classical Mechanics and Integrable Dynamical Systems. The research
of Yu. F. was supported by the MICINN-FEDER grant MTM2009-06973
and CUR-DIUE grant 2009SGR859.

\

Yuri N. Fedorov

Department de Matem\`atica I

Universitat Politecnica de Catalunya

Barcelona, E-08028 Spain

e-mail: Yuri.Fedorov@upc.es

\

Bo\v zidar Jovanovi\' c

Mathematical Institute SANU

Serbian Academy of Sciences and Arts

Kneza Mihaila 36, 11000, Belgrad, Serbia

e-mail: bozaj@mi.sanu.ac.yu


\begin{thebibliography}{99}

\small

\bibitem{AHP}
{\it M. R. Adams, J. Harnad, E. Previato}, Isospectral Hamiltonian
flows in finite and infinite dimensions. I. Generalized Moser
systems and moment maps into loop algebras. Comm. Math. Phys. {\bf 117}
(1988), no. 3, 451--500.

\bibitem{AHH}
{\it M. R. Adams, J. Harnad, J. Hurtubise}, Isospectral Hamiltonian
flows in finite and infinite dimensions. II. Integration of flows.
Comm. Math. Phys. {\bf 134} (1990), no. 3, 555--585.

\bibitem{AKN}  {\it V. I. Arnold, V. V. Kozlov, A. I. Neishtadt},
Mathematical aspects of classical and celestial mechanics. Itogi
Nauki i Tekhniki.  Sovr. Probl. Mat. Fundamental'nye
Napravleniya, Vol. {\bf 3}, VINITI, Moscow 1985 (Russian). English transl.:
Encyclopadia of Math. Sciences, Vol.{\bf 3}, Springer-Verlag, Berlin 1989.

\bibitem{ADN}
{\it A. Arvanitoyeorgos, V.V. Dzhepko, Yu. G. Nikonorov},
Invariant Einstein metrics on some homogeneous spaces of classical
Lie groups, Canadian Journal of Mathematics  {\bf 61}  (2009),
no. 6, 1201--1213,  arXiv: math/0612504 [math.DG].

\bibitem{Be} {\it A. Besse}, {\it Einstein Manifolds}, Springer, A Series of Modern Surveys in Mathematics, 1987.

\bibitem{BCMR} {\it A. M. Bloch, P. E. Crouch, J. E. Marsden, T. S. Ratiu},
The symmetric representation of the rigid body equations and their
discretization, Nonlinearity, {\bf 15}( 2002), 1309-1341.

\bibitem{BCS} {\it A. M. Bloch, P. E. Crouch, A. K. Sanyal},  A variational problem on Stiefel manifolds,
Nonlinearity, {\bf 19} (2006), 2247-2276.

\bibitem{Bog} {\it O. I. Bogoyavlenski},  New integrable problem of classical mechanics.
{\it Comm. Math. Phys.\/} {\bf 94} (1984), 255--269

\bibitem{Bo} {\it A. V. Bolsinov},  {Compatible Poisson brackets on Lie
algebras and the completeness of families of functions in
involution}, Izv. Acad. Nauk SSSR, Ser. matem. {\bf 55} (1991), No.1,
68-92  (Russian); English translation: Math. USSR-Izv., {\bf
38} (1992),  No.1, 69-90.

\bibitem{BJ1} {\it A. V. Bolsinov, B. Jovanovi\' c},
{Integrable geodesic flows on homogeneous spaces}.
{\it Matem. Sbornik} {\bf 192} (2001) no. 7, 21-40 (Russian);
English translation: {\it Sb. Mat.} {\bf 192} (2001), no. 7--8, 951--968.

\bibitem{BJ2} {\it A. V. Bolsinov, B. Jovanovi\' c},
Non-commutative integrability, moment map and geodesic flows.
{\it Annals of Global Analysis and Geometry} {\bf 23} (2003), no. 4, 305-322,
arXiv: math-ph/0109031.

\bibitem{BJ3} {\it A. V. Bolsinov, B. Jovanovi\' c},
Complete involutive algebras of functions on  cotangent bundles of
homogeneous spaces, Mathematische Zeitschrift
{\bf 246} (2004),  no.  1-2, 213--236.

\bibitem{BJ4} {\it A. V. Bolsinov, B. Jovanovi\' c},
Magnetic Flows on Homogeneous Spaces, Com. Mat. Helv., \textbf{83}
(2008), no. 3, 679�700, arXiv: math-ph/0609005.

\bibitem{Br}  {\it A. V. Brailov},
{Construction of complete integrable geodesic flows on compact
symmetric spaces.} {Izv. Acad. Nauk SSSR, Ser. matem.} {\bf
50} (1986), no.2, 661--674,  (Russian); English translation: {Math. USSR-Izv.} {\bf 50} (1986), No.4, 19--31.

\bibitem{Chasles} {\it M. Chasles},
Les lignes g\'eod\'esiques et les lignes de courbure des surfaces du segond degr\'e. {Journ. de Math.}
{\bf 11} (1846), 5--20.

\bibitem{Dirac} {\it P. A. Dirac},
On generalized Hamiltonian dynamics. { Can. J. Math.}
{\bf 2} (1950), no.2, 129--148.

\bibitem{DGJ} {\it V. Dragovi\' c, B. Gaji\' c, B. Jovanovi\' c},
Singular Manakov Flows and Geodesic Flows on Homogeneous Spaces,
Transform. Groups  {\bf 14}  (2009),  no. 3, 513--530,
arXiv:0901.2444 [math-ph]

\bibitem{DKN} {\it B. A. Dubrovin, I. M. Krichever, S. P. Novikov}, Integrable Systems I,
Itogi Nauki i Tekhniki. Sovr.Probl.Mat. Fund.Naprav. Vol.{\bf 4},
 VINITI, Moscow 1985 (Russian). English transl.:{  Encyclopaedia of Math.Sciences},
 Vol. {\bf 4}, 173-280, Springer-Verlag, Berlin 1989.

\bibitem{Ef} {\it D. I. Efimov},  The magnetic geodesic flows on a homogeneous symplectic manifold.
Siberian Mathematical Journal {\bf 46} (2005), no.1, 83-93.

\bibitem{Fe}
{\it Yu. N. Fedorov}, Integrable systems, Lax representation and
confocal quadrics, Amer. Math. Soc. Transl. (2) Vol. {\bf 168}, 173-199 (1995).

\bibitem{Fe2}
{\it Yu. N. Fedorov},  Integrable flows and Backlund transformations on extended Stiefel varieies with
application to the Euler top on the Lie group $SO(3)$, J.Non. Math. Phys. {\bf 12} (2005), Suppl. 2, 77-94.

\bibitem{FeJo}
{\it Yu. N. Fedorov, B. Jovanovi\' c},
Nonholonomic  LR systems as  Generalized Chaplygin systems with an
Invariant Measure and Geodesic Flows on Homogeneous Spaces, J.
Nonlinear Sci.  {\bf 14}  (2004),  no. 4, 341--381.  , arXiv:
math-ph/0307016

\bibitem{FeJo2}{\it Yu. N. Fedorov, B. Jovanovi\' c},
Continuous and discrete Neumann systems on Stiefel varieties as
matrix generalizations of the Jacobi--Mumford systems, in
preparation.

\bibitem{Fl}
{\it H. Flaschka}, Towards an algebro-geometric interpretation of the
Neumann system. Tohoku Math. J. (2) {\bf 36} (1984), no. 3, 407--426.

\bibitem{Griffits} {\it F. Griffits, J. Harris}, Principles of Algebraic Geometry. Wiley Interscience, New York 1978.

\bibitem{GS} {\it V. Guillemin, S.  Sternberg},
{\it Symplectic techniques in physics.} Cambrige University press, 1984.

\bibitem{Je} {\it G. Jensen}, Einstein metrics on principal
fiber bundles, J. Diff. Geom. {\bf 8} (1973), 599-614.

\bibitem{Kap89} {\it S. Kapustin},
The Neumann system on Stiefel varieties. Preprint 1992 (Russian).


\bibitem{KN} {\it S. Kobayashi, K.  Nomizu}, {\it Foundation of Differential Geometry},
Volume II. John Willey \& Sons, New York, 1969, 468 p.


\bibitem{Knorr1}  {\it H. Kn\"orrer},
Geodesics on quadrics and a mechanical problem of
C.Neumann. {J. Reine Angew. Math.} {\bf 334} (1982), 69--78.


\bibitem{Ku} {\it M. Kummer}, On the construction of the reduced
phase space of a Hamiltonian system with symmetry. Indiana Univ. Math. J. {\bf 30} (1981), 281-291.

\bibitem{Kuz}
{\it V. B. Kuznetsov},
Isomorphism of an $n$-dimensional Neumann system and an $n$-site Gaudin magnet. (Russian) Funktsional. Anal. i
Prilozhen. {\bf 26} (1992), no. 4, 88--90 (Russian); translation in Funct. Anal. Appl. {\bf 26} (1992), no. 4, 302--304.

\bibitem{Ma} {\it S. V. Manakov},  Note on the integrability of the Euler equations
of $n$--dimensional rigid body dynamics, Funkts. Anal. Prilozh. {\bf 10} (1976), no. 4, 93-94
(Russian).

\bibitem{Mik} {\it I. V. Mikityuk},
{Integrability of the Euler equations associated with filtrations
of semisimple Lie algebras}. {Matem. Sbornik} {\bf 125(167)} No.4
(1984) (Russian); English translation: {Math. USSR Sbornik} {\bf
53} (1986), No.2, 541-549.


\bibitem{Moser} {\it J. Moser},
Geometry of quadric and spectral theory. In: Chern Symposium 1979,
Berlin--Heidelberg--New York, 147--188, 1980.

\bibitem{Moser2}
{\it J. Moser}, Various aspects of integrable Hamiltonian systems.
Dynamical systems (C.I.M.E. Summer School, Bressanone, 1978), pp.
233--289, Progr. Math., 8, Birkh�user, Boston, Mass., 1980.

 \bibitem{Mum} {\it D. Mumford},
Tata lectures on theta, Birkha\"user, Boston 1984.

\bibitem{MF2}  {\it A. S. Mishchenko, A. T.  Fomenko},
{Generalized Liouville method of integration of Hamiltonian systems},
Funkts. Anal. Prilozh. {\bf 12} (1978), No.2, 46-56  (Russian);
English translation: Funct. Anal. Appl. {\bf 12}  (1978), 113-121.


 \bibitem{N} {\it N. N. Nekhoroshev},
{Action-angle variables and their generalization},
{Tr. Mosk. Mat. O.-va.} {\bf 26} (1972), 181-198,  (Russian);
English translation: {Trans. Mosc. Math. Soc.} {\bf 26} (1972),
180-198.

\bibitem{Neum} {\it  C. Neumann}, De probleme quodam mechanico, quod ad primam integralium
ultra-ellipticoram classem revocatum. {J. Reine Angew. Math.\/} {\bf 56} (1859).


\bibitem{Pe} {\it A. M. Perelomov},
Some remarks on the integrability of the equations of motion of a rigid body
in an ideal fluid, Funkt. Anal. Prilozh {\bf 15} (1981), no. 2, 83-85 (Russian);
English translation: Funct. Anal. Appl. {\bf 15} (1981), 144-146.

\bibitem{Ra} {\it T. Ratiu},  The C. Neumann problem as a completely integrable
system on an adjoint orbit, Trans. Amer. Math. Soc. {\bf 264} (1981), no.2, 321-329.

\bibitem{R} {\it A. G. Reyman},  Integrable
Hamiltonian systems connected with graded Lie algebras,
Zap. Nauchn. Semin. LOMI AN SSSR {\bf 95} (1980), 3-54 (Russian);
English translation: J. Sov. Math. {\bf 19} (1982), 1507-1545.


\bibitem{RS}
{\it A. G. Reyman, M. A.  Semonov-Tian-Shanski},
Group theoretical methods in the theory of finite dimensional integrable systems.
In. {Dynamical systems VII}  Itogi Nauki i Tekhniki. Sovr.Probl.Mat. Fund.Naprav. Vol.{\bf 16},
 VINITI, Moscow 1987 (Russian). English transl.:
Encyclopaedia of Math.Sciences, Vol. {\bf 16}, Springer 1994.

\bibitem{Sa}
{\it P. Saksida},  Nahm's Equations and Generalizations of the Neumann System,
Proc. London Math. Soc. {\bf 78} (1999), no. 3, 701-720.

\bibitem{Sa2} {\it P. Saksida},  Integrable anharmonic oscilators on spheres and hyperbolic spaces.
Nonlinearity {\bf 14} (2001), 977-994.

\bibitem{Su}
{\it Yu. B. Suris}, The problem of integrable discretization: Hamiltonian approach.
Progress in Mathematics, {\bf 219}. Birkh�user Verlag, Basel, 2003.

\bibitem{Th} {\it A. Thimm},
{Integrable geodesic flows on homogeneous spaces}
{\it Ergod. Th. \& Dynam. Sys.} {\bf 1} (1981), 495-517.

\bibitem{Ve} {\it A. P. Veselov},
Finite zone potentials and integrable systems on a sphere  with quadratic potential.
Funkt. Anal. Prilozh. {\bf 14} (1980), no. 1, 48-50 (Russian);
English translation: Funct. Anal. Appl. {\bf 14} 37-39 (1980).

\bibitem{Wo} {\it S. Wojciechowski},  Integrable one-partical potentials
related to the Neumann system and the Jacobi problem of geodesic
motion on an ellipsoid, Phys. Lett. A {\bf 107}  (1985), 107-111.

\bibitem{Zu} {\it N. T. Zung},  Torus actions and integrable systems.
In: Bolsinov A.V., Fomenko A.T., Oshemkov A.A. (eds.) {Topological
methods in the theory of integrable systems}. Cambridge Scientific
Publ., 2006, 289-328; arXiv: math.DS/0407455.

\end{thebibliography}
\end{document}